\documentclass[aps,prb,reprint,superscriptaddress,nofootinbib]{revtex4-2}

\usepackage{amsmath,amssymb,bm}
\usepackage{graphicx}
\usepackage{dcolumn}
\usepackage{booktabs}
\usepackage{xcolor}
\usepackage{hyperref}
\usepackage{siunitx}

\begin{document}

\title{Gradient Quantum Matter at Frustrated Interfaces}

\author{ Kaize Wang}
\affiliation{Max Planck Institute for the Structure and Dynamics of Matter, Hamburg, Germany}
\author{Sophie Rodehutskors}
\affiliation{Max Planck Institute for the Structure and Dynamics of Matter, Hamburg, Germany}
\author{Philip J. W. Moll}
\affiliation{Max Planck Institute for the Structure and Dynamics of Matter, Hamburg, Germany}
\author{J{\"o}rg Schmalian}
\affiliation{Institute for Theory of Condensed Matter, Karlsruhe Institute of Technology, 76131 Karlsruhe, Germany}
\affiliation{Institute for Quantum Materials and Technologies, Karlsruhe Institute of Technology, 76131 Karlsruhe, Germany}
\date{\today}

\begin{abstract}
 We develop a theoretical framework for using controlled strain gradients to create interfaces at which the frustration between competing electronic states stabilizes order absent from the surrounding bulk. 
While uniform strain often selects between symmetry-related electronic states,  strain gradients with a sign-changing strain field force an interpolation between them within a single crystal. 
 When  strain couples to a bilinear of a multi-component order parameter, the interpolation between strain-selected domains can activate additional ordering channels near the strain-neutral plane. For stripe magnetism, we obtain localized double-$Q$ textures with charge-order or spin-vorticity order, depending on the relative spin orientation. For two-component nematic superconductors, the two-dimensional interface can spontaneously break time-reversal symmetry while both bulk domains remain time-reversal symmetric.  We identify distinct mechanisms in tetragonal and trigonal crystals, including 
a separate interfacial axial-ordering transition
in the trigonal case. Piezomagnetic coupling extends the same principle to altermagnets, where a magnetic field converts the strain profile into a conjugate field for magnetic order. These results establish the concept of gradient quantum matter, in which bending-induced strain gradients create two-dimensional interfaces that support new states of two-dimensional order
\end{abstract}

\maketitle

\section{Introduction}
\label{sec:introduction}

Macroscopic degeneracies of quantum states are at the heart of the unconventional phases of matter commonly found in quantum materials. A central strategy in this field is to tune materials into regimes where competing energy scales become comparable or kinetic energy is quenched by frustration. The vast entropy, in the absence of a defined dominant ordering mechanism, allows sub-dominant orders to stabilize in large regions of the phase diagram. Countless examples include doping a Mott insulator in the cuprates~\cite{Emery1995,Lee2006,Keimer2015}, suppressing static nematic and magnetic order in the pnictides~\cite{Fernandes2022}, or balancing Kondo and RKKY scales in heavy fermions~\cite{Doniach1977}. Tuning parameters are either chemical substitution, such as carrier doping or chemical pressure, or environmental parameters such as magnetic fields, pressure or uniaxial strain. The latter are often considered clean as they are not accompanied by microscopic disorder. However, the divergent susceptibility at critical points strongly amplifies any spatial inhomogeneities and thus substantial efforts are made to ensure homogeneous tuning. For example, strain cells have been meticulously designed to deliver as homogeneous uniaxial strain to crystals as possible~\cite{Hicks2025}. Homogeneous strain was then used to efficiently tune quantum materials and their symmetry~\cite{Fisher2011,Moll2018,Ikeda2018,Jo2024}.

\begin{figure}[tb]
    \centering
    \includegraphics[width=\linewidth]{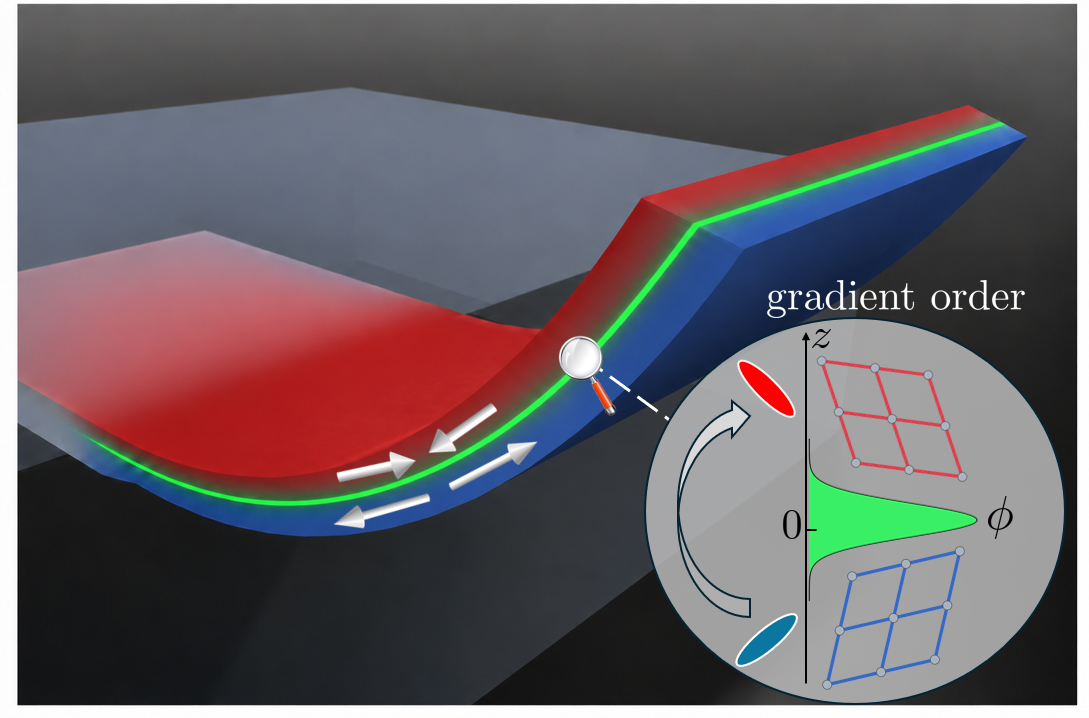}
    \caption{Schematic for bending-induced gradient order. Red and blue denote distinct electronic orders favored by tensile and compressive strain on opposite sides of the strain-neutral plane. Frustration between these strain-selected domains stabilizes an additional interfacial order, shown in green. White arrows indicate stretching and compression of the corresponding layers. The inset depicts the opposite strain-induced rhombic lattice distortions in the two domains and the gradient-order amplitude $\phi(z)$, localized near the neutral plane at $z=0$, where $z$ measures distance along the local thickness direction.}
    \label{fig:generalidea}
\end{figure}

Here, we introduce the  paradigm in which new collective states emerge explicitly from strong, controlled gradients of tuning parameters that are inaccessible to homogeneous tuning, which we refer to as {\em gradient quantum matter}. The general concept is that gradients allow a tuning parameter to change sign within a crystal, creating a unique environment around the zero crossing of the tuning parameter. There, the material is unbiased, yet strongly frustrated due to the vicinity of oppositely biased material on either side. This frustration destabilizes a primary order parameter that is sensitive to the tuning, and the resulting emergent degrees of freedom may form a new type of gradient order. While this general concept will find applications in various quantum materials where strong gradients in conjugate fields can be achieved, here we focus specifically on the case of strain-tuning. As illustrated schematically in Fig.~\ref{fig:generalidea}, a sign-changing strain caused by bending can result in distinct electronic configurations in the tensile and compressive layers, whose frustrated interpolation can stabilize additional gradient order near the strain-neutral plane.

The effects of homogeneous bending are particularly direct when a collective degree of freedom couples bi-linearly to a component of the strain tensor. Schematically, if $N$ denotes an order parameter or a composite of  order-parameter components and $\varepsilon_\Gamma$ a strain component transforming in the same irreducible representation $\Gamma$ of the crystal point group, then the free energy contains the term
\begin{equation}
    F_{\mathrm{int}}=-\lambda\int d^d r\,\varepsilon_\Gamma(\bm r)N(\bm r).
    \label{eq:bilinear_coupling}
\end{equation}
Thus, spatially varying strain acts as a position-dependent conjugate field for $N$. It is therefore sensitive not only to the magnitude, but also to the sign of strain. A bent sample can consequently force the order parameter to change in magnitude and sign across the system, creating controlled textures, interfaces, and domain structures. 
This mechanism is distinct from flexo-type effects such as flexo-electricity~\cite{Zubko2013,Yudin2013,Wang2019,Tang2025,Edstrom2026}, in which a spatial strain gradient induces an electric polarization. Such couplings enter at higher order in spatial derivatives, often rely on the inversion-symmetry breaking associated with strain gradients, and can generate a response even when the strain does not change sign. By contrast, the key ingredient in Eq.~\eqref{eq:bilinear_coupling} is the frustration of the order parameter at an interface where the strain changes sign. This frustration can stabilize additional interfacial order and strongly enhance the response to an imposed strain gradient. Indeed, the manipulation of boundary conditions through strain engineering was recently used to probe nonreciprocal charge transport in the iron-based superconductor ${\mathrm{FeTe}}_{0.7}{\mathrm{Se}}_{0.3}$~\cite{Li2026} and in the topological insulator Bi$_2$Se$_3$~\cite{Veremchuk2026}.

Electronic nematic order~\cite{Kivelson1998,Fernandes2010,Chu2012,Fernandes2014,Fradkin2015,Fernandes2019,Fernandes2016} provides a paradigmatic example of the coupling in Eq.~\eqref{eq:bilinear_coupling}, with $N$ denoting the nematic order parameter. Like the corresponding shear strain, this order parameter transforms non-trivially under point-group rotations. Their identical transformation properties permit the linear coupling in Eq.~\eqref{eq:bilinear_coupling}. Bending therefore provides a natural way to impose a sign-changing field conjugate to nematic order and to probe the response of a nematic quantum material across an engineered domain wall.

A second class of examples is provided by altermagnetic order~\cite{Smejkal2022,Smejkal2022_2} in an external magnetic field. Piezomagnetic couplings~\cite{Patri2019,Sorensen2021,Steward2023,Aoyama2024,McClarty2024,Chakraborty2025,Takahashi2025,Ohlendorf2026} allow combinations of strain and magnetic field to act as fields conjugate to components of the altermagnetic order parameter. In this case, $N$ in Eq.~\eqref{eq:bilinear_coupling} is proportional to the product of the altermagnetic order parameter and the magnetic field. Bending in a magnetic field can therefore generate a spatially varying conjugate field for altermagnetism, offering a route to the controlled formation of altermagnetic textures even when strain alone cannot couple linearly to the magnetic order parameter.

The situation becomes especially rich for ordered states described by a multicomponent
order parameter that form composites that transform like strain~\cite{Fradkin2015,Fernandes2019}.   In this case $N$ is not a primary microscopic order parameter but is  built from more fundamental  degrees of freedom. Examples include nematic order generated by magnetic or charge-density-wave fluctuations or composite orders associated with multi-component superconductivity. A strain field then couples directly to the composite variable while indirectly reorganizing its underlying constituents. Spatially varying strain can consequently expose information that is  impossible to obtain from any homogeneous symmetry-breaking field. On the one hand, the internal structure of the multi-component order can manifest itself in characteristic length scales, domain-wall profiles, defect structures, induced secondary order, and nonlinear response. In addition, new symmetry breaking -- absent in the homogeneous bulk -- may form and allow shaping states such as spin vortex crystals or time-reversal symmetry breaking superconductivity. 

Experimental routes to engineering spatially varying strain include bent crystalline cantilevers fabricated by focused-ion-beam machining~\cite{Diaz2022}, freestanding crystalline membranes~\cite{Wang2026}, and lattice relaxation gradients in epitaxial films~\cite{Catalan2005}. Our mechanism requires the relevant strain-induced conjugate field to change sign. We therefore focus on controlled bending as a concrete realization, where tensile and compressive regions are naturally separated by a strain-neutral plane. The detailed geometry and simulation are discussed in  Appendix~\ref{app:fem}.

In this paper, we develop a Ginzburg–Landau framework for electronic interfaces created by a sign-changing strain field. We first establish how bending selects and pins domain walls in scalar nematic order and, through piezomagnetic coupling in an external magnetic field, in altermagnetic order. We then show that multi-component ordered states possess additional degrees of freedom that can become active within these walls. For stripe magnetism, the overlap of orthogonal ordering wave-vectors produces a localized double-$Q$ texture with either a checkerboard charge-order bilinear or spin vorticity, depending on the interaction that selects the relative spin orientation. We derive analytically and confirm numerically profiles that relate the width of this interfacial order to the magnetic stiffness and competing-order anisotropy. For nematic superconductors, we identify two distinct routes to interfacial time-reversal symmetry breaking. In tetragonal crystals, a quartic coupling favoring a complex relative phase
makes an interpolation with axial order energetically preferable
whenever the two components overlap, although the homogeneous ground states remain nematic. In trigonal crystals, crystalline pinning can instead drive a separate interfacial transition by overcoming the energetic cost of time-reversal symmetry breaking. We obtain an analytical instability criterion and the spatial profile of the emerging axial mode. We also discuss extensions to charge-density-wave and vestigial order. Together, these results establish how controlled bending can reveal competing electronic configurations through order localized at an internal interface.

\section{Geometry and elasticity of mesoscopic bending}
\label{sec:geometry}

We consider an initially straight crystalline strip of length $D$ along $x$, width $w$ along $y$, and thickness $L$ along $z$, with $D\gg w\gg L$, bent about the $y$ axis as illustrated in Fig.~\ref{fig:bendingschematic}. In the bent configuration, \(z\in[-L/2,L/2]\) denotes the signed distance from the neutral surface measured along its local normal. Let $s$ denote arc length along the neutral line and $\theta(s)$ its tangent angle. Within Euler--Bernoulli beam theory~\cite{LandauLifshitzElasticity1986}, the curvature $\kappa(s)=-d\theta/ds$ produces a longitudinal strain $\varepsilon_{xx}(s,z)=z\kappa(s)$.

The strain changes sign across $z=0$, separating tensile and compressive regions. Approximately uniform curvature therefore generates a strain profile that is linear across the thickness and nearly uniform along the strip. A cantilever implementation and a finite-element analysis of its strain distribution, both confirming such a strain profile,  are presented in Appendix~\ref{app:fem}.

We consider strain profiles that vary smoothly on atomic scales, allowing a continuum description of the electronic response. Percent-level strain variations over micrometer distances have been demonstrated in bent crystalline micro-cantilevers~\cite{Diaz2022}.
Although bending produces an approximately linear strain profile across the thickness, we use a bounded parametrization to impose well-defined bulk boundary conditions:
\begin{equation}
\varepsilon_\Gamma(z)
=
\varepsilon_\infty\tanh\left(\frac{z}{L}\right).
\label{eq:tanh_profile}
\end{equation}
Here $\varepsilon_\Gamma$ is the relevant symmetry-resolved strain component, $\varepsilon_\infty$ sets its asymptotic magnitude, and $L$ sets the strain-variation length. Near the neutral plane, this profile reproduces the linear dependence $\varepsilon_\Gamma(z)\simeq\varepsilon_\infty z/L$, while approaching opposite constant values away from the interface. Eq.~\eqref{eq:tanh_profile}  is not intended to represent a  microscopic elastic solution. Rather, it captures the generic situation in which the strain saturates away from a crossover region and changes sign smoothly over a mesoscopic distance $\sim L$.

\section{Strain gradients for scalar nematic and altermagnetic order}
\label{sec:bilinear}
\begin{figure}
    \centering
    \includegraphics[width=\linewidth]{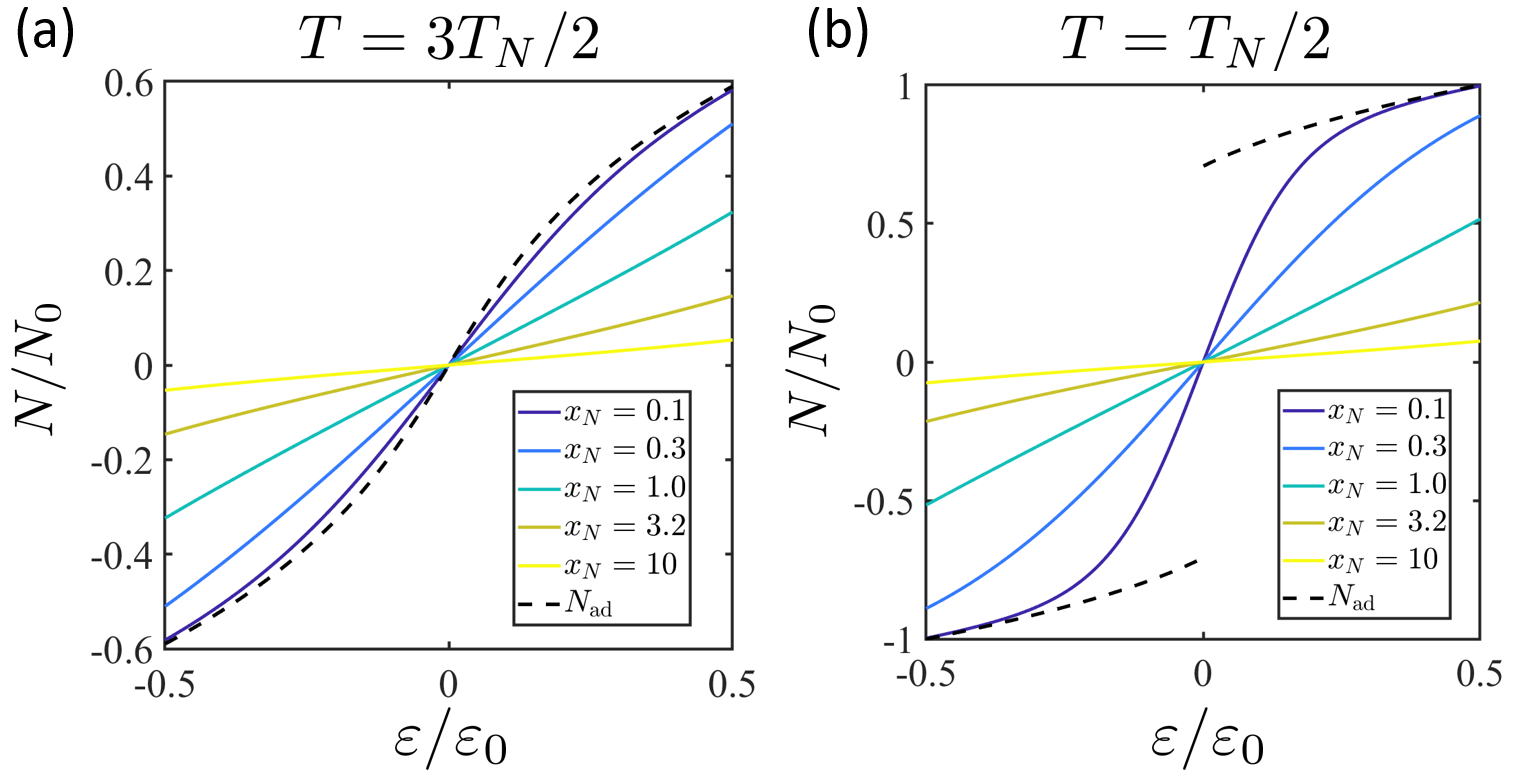}
    \caption{Scalar nematic response to sign-changing strain at (a) $T=3T_N/2$ and (b) $T=T_N/2$. Solid curves solve the full GL equation for $x_N=\sqrt{K_N/(a_0T_N L^2)}=0.1,0.3,1,3.2,10$; black dashed curves show the local adiabatic equilibrium response $N_\mathrm{ad}$ at $K_N=0$. The profiles are plotted against the local strain $\varepsilon(z)$, with $N_0=\sqrt{a_0T_N/u}$ and $\varepsilon_0=(a_0T_N)^{3/2}/(|\lambda|\sqrt u)$.}
    \label{fig:scalar_nematic_response}
\end{figure}
To set the stage, we first describe nematicity by a single real scalar field $N$, which changes sign under a $\pi/2$ in-plane rotation and transforms in the same symmetry channel as the strain component $\varepsilon_{B_{1g}}\left(z\right)=\varepsilon_{xx}\left(z\right)-\varepsilon_{yy}\left(z\right)$. Strain therefore acts as a conjugate field of $N(z)$. The minimal Ginzburg--Landau functional is
\begin{equation}
 \mathcal{F}=\int d^d r\,
 \left[\frac{K_{N}}{2}(\nabla N)^2
       +\frac{a}{2}N^2+\frac{u}{4}N^4
       -\lambda\varepsilon_{B_{1g}}(\mathbf r)N\right],
 \label{eq:scalar_nematic_gl}
\end{equation}
where $K_{N},u>0$, $a=a_0(T-T_{N})$ with $a_0>0$ and $T_N$ the nematic transition temperature, and $\lambda$ is the nemato-elastic coupling. For the bending profile $\varepsilon_{B_{1g}}(z)$ such as the one of Eq.~\eqref{eq:tanh_profile}, minimization gives
\begin{equation}
 -K_{N}\frac{d^2N}{dz^2}+aN+uN^3
 =\lambda\varepsilon_{B_{1g}}(z).
 \label{eq:scalar_nematic_el}
\end{equation}

Fig.~\ref{fig:scalar_nematic_response} compares the full solution of this non-linear differential equation with the local approximation $aN+uN^3=\lambda\varepsilon(z)$, valid for $K_N\rightarrow 0$. Above $T_{N}$, strain induces a finite nematic response, while the proper inclusion of the gradient term smoothens its spatial variation; see Fig.~\ref{fig:scalar_nematic_response}(a). Below $T_{N}$, the strain reversal favors opposite nematic domains. The local equilibrium branch switches discontinuously at $z=0$, whereas finite $K_{N}$ produces a continuous domain wall pinned to the neutral plane; see Fig.~\ref{fig:scalar_nematic_response}(b). Increasing $K_{N}/(a_0T_{N}L^2)$ increases the domain-wall width.
For $T<T_{N}$ and weak strain across the wall core, the leading profile is the well-known kink solution
\begin{equation}
 N(z)\simeq\sqrt{\frac{|a|}{u}}\,
 \tanh\!\left(\frac{z}{\sqrt{2}\,\xi_{N}}\right),
 \qquad \xi_{N}=\sqrt{\frac{K_{N}}{|a|}}.
 \label{eq:scalar_nematic_wall}
\end{equation}
The strain profile thus selects and pins the domains, while the nematic stiffness sets the intrinsic wall width.

These considerations can be extended to magnetic order
parameters, although an important modification is required. Strain is even
under time reversal and therefore cannot couple bilinearly to a
time-reversal-odd magnetic order parameter. In altermagnets, however, the
magnetic order generally transforms non-trivially under crystalline
symmetries and can be viewed as a higher-rank multi-polar magnetic order parameter~\cite{Smejkal2022}. This allows strain to couple linearly to the altermagnetic order
in the presence of an external magnetic field. The resulting interaction is
the Landau-theory manifestation of piezomagnetism~\cite{Steward2023,Aoyama2024,McClarty2024,Chakraborty2025,Takahashi2025,Ohlendorf2026}.

To see this, let $M_{\rm AM}$ denote an altermagnetic order parameter
transforming according to an irreducible representation $\Gamma_{\rm AM}^{-}$,
where the superscript indicates that $M_{\rm AM}$ is odd under time reversal.
A symmetry-adapted strain component $\varepsilon_{\Gamma}$ is
time-reversal even, while a magnetic-field component $H_\alpha$ transforms
according to the  time-reversal-odd representation
$\Gamma_{J_\alpha}^{-}$ of a magnetic dipole. The
combination $ N \sim H_\alpha M_{\rm AM}$
is therefore time-reversal even and, provided 
\begin{equation}
    \Gamma
    \subset
    \Gamma_{J_\alpha}^{-}
    \otimes
    \Gamma_{\rm AM}^{-},
    \label{eq:alt_selection_rule}
\end{equation}
it couples to strain $\varepsilon_\Gamma$ according to Eq.~\eqref{eq:bilinear_coupling}.
Hence, in an altermagnet,
the external magnetic field supplies precisely the additional
time-reversal-odd factor needed to couple to strain. 
The tensile and compressive parts of the bent crystal therefore favor
opposite altermagnetic domains. Below the altermagnetic transition this
provides a direct way of generating and pinning an altermagnetic domain
wall close to the neutral plane, complementary to approaches based on inhomogeneous magnetic fields~\cite{Amin2024A,Gomonay2024}.
Reversing the magnetic field interchanges
the preferred domains, whereas rotating the magnetic field can select a
different altermagnetic symmetry channel altogether.
Such strain-engineered domains, also discussed recently in Ref.~\cite{Melendez2026}, should  provide an interesting setting
for magnetotransport. Altermagnetic order produces strongly
momentum-dependent spin splitting even in the absence of a uniform
magnetization, and different altermagnetic domains correspond to
symmetry-related but distinct spin-split electronic structures. A bent
device containing two oppositely selected domains and a controlled
interface between them can consequently exhibit domain-dependent
longitudinal and transverse transport.

\section{Strain-coupled order parameter bilinears and interfacial orders}
The preceding section considered a scalar nematic order parameter to which strain couples directly as a conjugate field. We now consider multi-component ordered states, where the strain-coupled variable is a bilinear of degenerate primary fields, i.e. symmetry dictated degenerate multi-component order parameters~\cite{Ozaki1986,SigristUeda1991,Fradkin2015,Hecker2018,Fernandes2019,Halcrow2025}. In this setting, strain selects the internal configuration of the order, while its spatial variation can create interfaces with properties distinct from those of the neighboring bulk domains.

Let $\boldsymbol{\eta}$ denote a primary order parameter with $n_\eta>1$ symmetry-related components, i.e. the order parameter transforms under some higher-dimensional irreducible representation of the symmetry group of the Hamiltonian. Bilinears of the form
\begin{equation}
    B=\boldsymbol{\eta}^{\dagger}{\cal B} \boldsymbol{\eta}
    \label{eq:bilinears_general}
\end{equation}
may transform in different symmetry channels, where ${\cal B}$ is an $n_\eta\times n_\eta $ dimensional matrix. For the slowly varying strain considered here, a linear strain coupling Eq.~\eqref{eq:bilinear_coupling} is allowed when the bilinear 
transforms in the same representation as the corresponding strain component, i.e. it is a composite nematic order parameter
\begin{equation}
    N=B_{\rm nem}=\boldsymbol{\eta}^{\dagger}{\cal B}_{\rm nem} \boldsymbol{\eta}.
\end{equation}
Strain therefore acts as a position-dependent anisotropy among the primary components, i.e. lifting at least partially the degeneracy of the higher-dimensional irreducible representation.  We focus in particular on the regime in which the primary order is established, i.e., below the transition temperature, so these bilinears characterize its amplitude and internal structure. Below we also comment on nearly-degenerate distinct single-component order parameters and on the regime of vestigial nematic order where in the strain-free bulk only the composite $N$ acquires a nonzero expectation value; see e.g. Ref.~\cite{Fernandes2019}.

When strain selects different configurations on opposite sides of the neutral plane, the gradient energy requires a continuous interpolation between them. This interpolation  involves changes in component amplitudes  or relative phases. Where different ordered states, that ordinarily exclude each other, occur in the same spatial region,  additional bilinears can become nonzero and produce localized interfacial order that is the consequence of the strain-imposed competition of degenerate states.

In Sec.~\ref{sec:stripe_magnetism}, we examine stripe magnetism~\cite{Huang2008,Zhao2008,Goldman2008} widely discussed in the context of iron-pnictides~\cite{Johnston2010,Hirschfeld2011,Dai2015,Fernandes2022}, formulate the direct strain coupling to its nematic bilinear, and determine how the magnetic components connect across the $C_4$-symmetric interface. Their relative spin orientation distinguishes collinear charge density wave double-$Q$ order~\cite{Fernandes2016,Fernandes2019,Bohmer2015,Allred2016} from a non-collinear spin-vortex pattern~\cite{Fernandes2016,Fernandes2019,Meier2018}. In Sec.~\ref{sec:cdw_vestigial}, we apply the same framework to bidirectional charge-density-wave order discussed in the cuprate superconductors~\cite{Hanaguri2004,Robertson2006,Wise2008,Ghiringhelli2012,Chang2012,BlancoCanosa2013,LeTacon2014,Comin2015,Zhou2025}, commenting on strain selection between symmetry-related ordering directions and the overlap of their complex amplitudes. Finally, in Sec.~\ref{sec:TRSBSC}, we study two-component nematic superconductors. We introduce their 
gauge-invariant nematic and axial bilinears and analyze the tetragonal and trigonal cases, showing how the relative-phase freedom can stabilize a time-reversal-breaking superconducting interface between time-reversal-symmetric bulk domains. In particular, the trigonal case may be relevant to doped Bi$_2$Se$_3$ where ample evidence exists in favor of nematic superconductivity~\cite{Matano2016,Yonezawa2017,Pan2016,Du2017,Asaba2017,Willa2018,Tao2018,Kostylev2020,Cho2020}.
Nematic superconductivity has also been discussed in the topological semimetal CaSn$_3$~\cite{Siddiquee2022}, where evidence has been reported in favor of a two-component $E_u$ pairing state of the cubic point group O$_h$. The associated nematic bilinear transforms according to the two-component $E_g$ representation. The theory developed in Sec.~\ref{sec:trigonal} can  be applied to this system essentially without modification.

The following sections consider two-component order parameters whose degeneracy is protected by symmetry, i.e. that belong to a two-dimensional irreducible representation of the symmetry group. The analysis can, however, be readily generalized to two nearly degenerate single-component order parameters. If their product representation transforms as a strain component, most of the arguments below carry over directly. One example is a pair of $d_{xy}$- and $g$-wave superconducting order parameters in a tetragonal system, for which the bilinear $\psi^*_d\psi_g+h.c.$ couples to $\varepsilon_{xx}-\varepsilon_{yy}$. Related work on nearly degenerate states strongly suggests that time-reversal symmetry breaking can also emerge in the neutral plane of such systems, provided the states are sufficiently close in energy~\cite{Willa2021,Yuan2021}.

\subsection{Stripe magnetism and interfacial double-$Q$ order}
\label{sec:stripe_magnetism}

In many iron-based superconductors, stripe magnetism is described by
two real vector order parameters, $\mathbf S_X$ and $\mathbf S_Y$,
associated with the symmetry-related ordering vectors
\begin{equation}
    \mathbf Q_X=\left(\frac{\pi}{a},0\right),
    \qquad
    \mathbf Q_Y=\left(0,\frac{\pi}{a}\right).
\end{equation}
For a bending-induced texture varying along $z$, the spin density is
\begin{equation}
    \mathbf S(\mathbf r)
    =\mathbf S_X(z)e^{i\mathbf Q_X\cdot\mathbf r}
    +\mathbf S_Y(z)e^{i\mathbf Q_Y\cdot\mathbf r}.
    \label{eq:spin_density}
\end{equation}
The two vectors ${\mathbf S_X}(z)$ and ${\mathbf S_Y}(z)$ make up the six-component order parameter $\mathbf{\eta}$ of Eq.~\eqref{eq:bilinears_general}. Important magnetic bilinears and their physical channels are listed in Table~\ref{tab:magnetic_bilinears}. The bilinear $R$ measures the total magnetic amplitude; a nonzero expectation value does not signal a broken symmetry. In contrast, $N$ distinguishes the stripe orientations and has $B_{1g}$ nematic symmetry, while the $C$ and $V$ are called the double-$Q$ orders~\cite{Fernandes2016,Fernandes2019,Bohmer2015,Allred2016,Meier2018}, carrying a wave vector $\mathbf{Q} =\mathbf{Q}_X+\mathbf{Q}_Y = (\pi/a,\pi/a)$. $C$ gives rise to a translational symmetry breaking checkerboard charge density wave order and $V$ corresponds to  inversion symmetry breaking spin-vorticity.

\begin{table}[tb]
 \centering
 \small
 \setlength{\tabcolsep}{5pt}
 \renewcommand{\arraystretch}{1.25}
 \begin{tabular}{lc}
  \hline\hline
  Bilinear & Physical channel \\
  \hline
  $R=\bm S_X^2+\bm S_Y^2$ & Magnetic amplitude \\
  $N=\bm S_X^2-\bm S_Y^2$ & $B_{1g}$ Nematic \\
  $V=(\bm S_X\times\bm S_Y)\cdot\hat{\bm z}$
      & Spin-vorticity \\
  $C=\bm S_X\cdot\bm S_Y$ & Checkerboard CDW \\
  \hline\hline
 \end{tabular}
  \caption{Time-reversal-even magnetic bilinears formed from the two components $\mathbf{S}_X$ and $\mathbf{S}_Y$ of the spin density of Eq.~\eqref{eq:spin_density}; see Ref.~\cite{Fernandes2016}.
In distinction to the symmetry-preserving $R$, similarly present in a single-component system, the bilinears $N$, $V$, and $C$ break additional symmetries and serve as additional order parameters. }
  \label{tab:magnetic_bilinears}
\end{table}

\begin{figure}[htb]
    \centering
    \includegraphics[width=\linewidth]{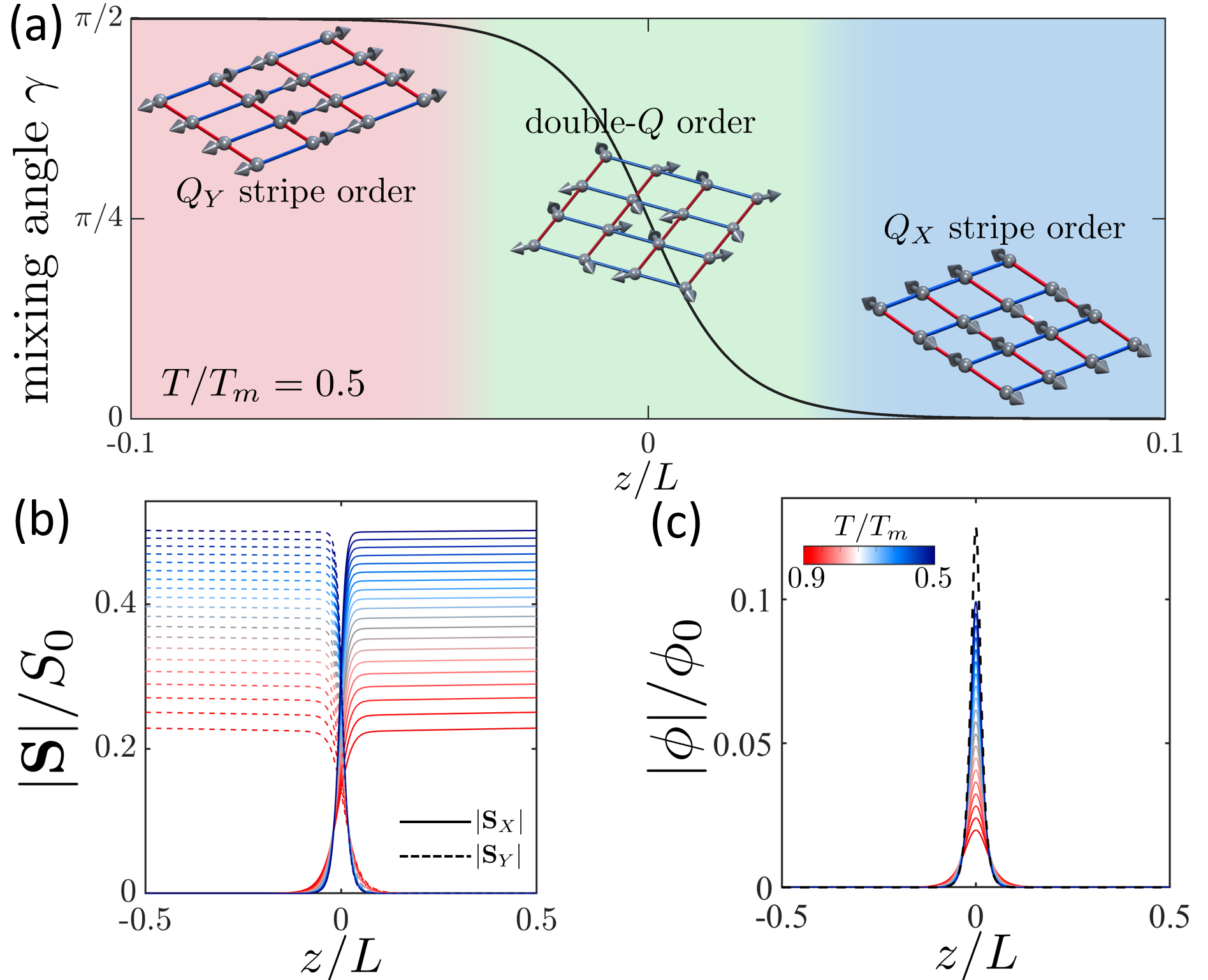}
    \caption{Bending-induced interfacial double-$Q$ order. (a) Mixing-angle profile $\gamma=\tan^{-1}(S_Y/S_X)$ at $T/T_m=0.5$.  Insets illustrate the interpolation between the $Q_Y$ and $Q_X$ stripe domains through an interfacial double-$Q$ texture. (b)Spatial amplitude profiles of magnetic order parameter $|\mathbf{S}_{X}|$(solid line) and $|\mathbf{S}_Y|$(dashed line) in units of $S_0=\sqrt{2a_mT_m/(u-g)}$ at different temperatures.  (c) Spatial amplitude profiles of the emergent double-$Q$ bilinear $|\phi(z)|$ in unit of $\phi_0 = S_0^2$. Colors in (b,c) indicate temperatures from $T/T_m=0.5$ (blue) to $0.9$ (red). Black dashed curves in (c) show the weak-strain, constant-amplitude analytical solution at $T/T_m=0.5$. The other GL parameters used here are given in Appendix~\ref{app:stripe_gl}.}
    \label{fig:phi_S}
\end{figure}

The nematic strain $\varepsilon_{B_{1g}}(z)=\varepsilon_{xx}(z)-\varepsilon_{yy}(z)$ couples directly to $N$. Its sign reversal selects opposite stripe orientations on the two sides of the neutral plane, as illustrated in Fig.~\ref{fig:phi_S}(a). The gradient energy requires a continuous interpolation between these domains, allowing the two magnetic components to overlap near the interface. This overlap supports a collinear double-$Q$ texture with finite $C$, or a spin-vorticity texture with finite $V$, depending on the relative spin orientation. We describe the texture using the following GL functional
\begin{equation}
\begin{aligned}
     f &= \frac{K_s}{2}(|\partial_z \mathbf{S}_X|^2 + |\partial_z \mathbf{S}_Y|^2) + \frac{r}{2} R + \frac{u}{4}R^2 \\
    &- \frac{g}{4} N^2 + w C^2 - h(z) N,
\end{aligned}
\end{equation}
where $h(z) = \lambda \varepsilon_{B_{1g}}(z)=h_\infty \tanh(z/L)$ combines the $z$-dependent $B_{1g}$ strain and the nemato-elastic coupling $\lambda$. For $w<0$ the system prefers collinear spins, making  $C$ and thus double-$Q$ CDW the sub-leading competing state, while for $w>0$  mutually perpendicular spins confined to the $xy$ plane are preferred, yielding $V$ and double-$Q$ spin-vortex crystals as main competitor. Depending on the sign of $w$ we define the  competing emergent bilinear order parameter 
\begin{equation}
\phi\left(z\right)=\left\{ \begin{array}{ccc}
C\left(z\right) & {\rm if} & w<0\\
V\left(z\right) & {\rm if} & w>0
\end{array}\right.
\label{eq:leading_competitor}
\end{equation}
Within the formalism of Eq.~\eqref{eq:bilinears_general}, $C$ and $V$ correspond to two bilinears that one can form from ${\mathbf S_X}(z)$ and ${\mathbf S_Y}(z)$. 
In Ref.~\cite{Hoyer2016} it was shown within a weak-coupling approach, that the sign of $w$ can be changed through disorder substitution.
Here, we take $K_s,\lambda>0$, $u>g>0$, and $w>-g$, ensuring quartic stability and a preference for single-$Q$ stripe order in the homogeneous bulk. The sign of the nemato-elastic coupling can be reversed through the mirror operation $z\rightarrow -z$.

The magnetic mass is $r=a_m(T-T_m)$, with $a_m>0$. We focus on the magnetically ordered regime $r<0$, where the unstrained system supports single-$Q$ stripe order. The sign-changing strain selects opposite stripe orientations in the top and bottom parts of the system, imposing the boundary conditions 
\begin{eqnarray}
    \left(\mathbf S_X(-L),\mathbf S_Y(-L)\right)&=&\left(0,S_\infty\hat{\mathbf{n}}_Y\right),\nonumber \\
    \left(\mathbf S_X(+L),\mathbf S_Y(+L)\right)&=&\left(S_\infty\hat{ \mathbf{n}}_X,0\right),
\end{eqnarray}
with $S_\infty=\sqrt{(2h_\infty-r)/(u-g)}$. The unit vectors $\hat{ \mathbf{n}}_X$ and $\hat{ \mathbf{n}}_Y$ specify the strain-selected  spin configuration on the bottom and top of the sample for $h_\infty>0$. Changing the sign of $h_\infty$ merely changes the configurations at $\pm\infty$.

We consider collinear spins, $\hat{\mathbf{n}}_ X \parallel \hat{\mathbf{n}}_Y$, for the $C$ double-$Q$ order, and perpendicular in-plane spins, $\hat{\mathbf{n}}_X \perp \hat{\mathbf{n}}_Y$, for the $V$ double-$Q$ order. For both, the analytical and numerical analysis it is convenient to use the strain profile $\varepsilon_\Gamma(z)$ of Eq.~\eqref{eq:tanh_profile} and extend the formal regime of $z$ to $\pm \infty$. The finite extend of the system along along the $z$-direction is contained in the strain profile of Eq.~\eqref{eq:tanh_profile}.

Minimizing the free energy with these strain-selected boundary conditions and keeping the spin axes $\hat{\mathbf{n}}_X$ and $\hat{\mathbf{n}}_Y$ fixed within each prescribed configuration give the spatial profiles of the primary magnetic order parameters $\mathbf{S}_X$ and $\mathbf{S}_Y$ amplitude in Fig.~\ref{fig:phi_S}(b). Writing $\mathbf{S}_X(z)=A(z)\cos\gamma(z)\,\hat{\mathbf{n}}_X$ and $\mathbf{S}_Y(z)=A(z)\sin\gamma(z)\,\hat{\mathbf{n}}_Y$, the solution in Fig.~\ref{fig:phi_S}(a) evolves smoothly from $\gamma=\pi/2$ in the $Q_Y$ domain to $\gamma=0$ in the $Q_X$ domain. Reflection symmetry gives $|\mathbf{S}_X(0)|=|\mathbf{S}_Y(0)|$ and hence $\gamma(0)=\pi/4$, identifying the neutral plane as the center of the double-$Q$ texture where the $D_{4h}$ tetragonal symmetry is recovered.

In either spin configuration, the emergent double-$Q$ bilinear amplitude is 
\begin{equation}
    |\phi|=|\mathbf{S}_X||\mathbf{S}_Y|.
\end{equation}
Fig.~\ref{fig:phi_S}(c) shows that it is localized near the neutral plane and vanishes in the single-$Q$ bulk domains. Cooling increases its peak and narrows its spatial profile.

\begin{figure}[tb]
    \centering
    \includegraphics[width=\linewidth]{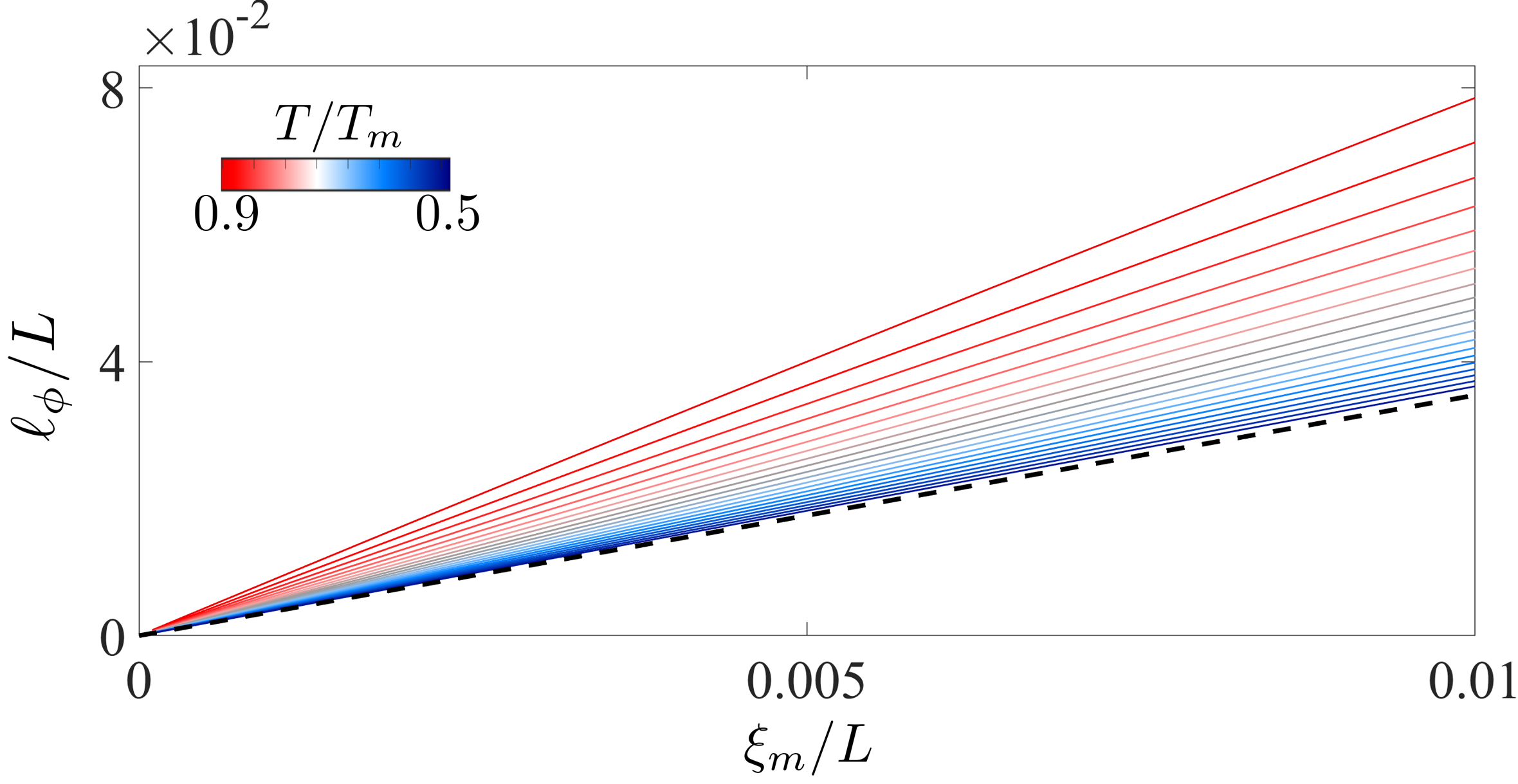}
    \caption{Spatial length scale $\ell_\phi$ of the double-$Q$ order versus bare magnetic correlation length $\xi_m/L$ at different temperature. The GL parameters are the same as in Fig.~\ref{fig:phi_S}. Black dashed line is the analytic approximation at $T=0.5 T_m$. }
    \label{fig:phi_width}
\end{figure}

 Fig.~\ref{fig:phi_width} quantifies this localization through the second-moment width
$
\ell_\phi^2=
 { \int dz\,z^2|\phi(z)|}/
{ \int dz\,|\phi(z)|}
$.
The numerically determined width grows approximately linearly with the bare magnetic length $\xi_m=\sqrt{K_s/(a_mT_m)}$ and increases on approaching $T_m$ within the temperature range shown. The numerical parameters and solution procedure are derived in  Appendix~\ref{app:stripe_gl}.

These numerical results admit a simple analytical description when the wall is narrow compared with $L$ and strain is weak across its core. In Appendix~\ref{app:stripe_gl}, we take $A$ as the  constant  bulk value  for weak strain limit and neglect the local strain inside the wall. Under these assumptions, we have 
\begin{equation}
    {|\phi(z)|} 
=\frac{a_m(T_m-T)}{2(u-g)}
\operatorname{sech}\!\left(\frac{\pi z}{2\ell_\phi}\right),\\
\end{equation}
where $\ell_\phi={\pi\xi_m}/{\sqrt{8(1-T/T_m)g_{\rm eff}/(u-g)}}$ is the characteristic length of the double-$Q$ order, with $g_{\rm eff}=g$ for $w>0$ and $g_{\rm eff}=g+w$ for $w<0$.
This analytic expression applies when $\ell_\phi\ll L$ and $h_\infty\ll a_m(T_m-T)$ at low temperature. It accounts for the growth of the peak on cooling and the nearly linear width dependence on $\xi_m$. The black dashed curves in Fig.~\ref{fig:phi_S}(c) and Fig.~\ref{fig:phi_width} show this result at $T/T_m=0.5$ which agrees well with the full numerical solution. To estimate the corresponding physical scale in real materials, we take a representative magnetic length $\xi_m\approx 1.6\mathrm{nm}$ measured in optimally doped $\mathrm{BaFe_{1.85}Co_{0.15}As_2}$\cite{Inosov2009}. For a bent flexure thickness $L=1\mu\mathrm{m}$ and $T=0.5T_m$, with the other GL parameters we used for Fig.~\ref{fig:phi_width}, the double-$Q$ width is around $5.6\mathrm{nm}$. 

In summary, inhomogeneous strain induced by bending can stabilize new
states near the neutral plane that are absent from the homogeneous bulk phase diagram. These interface orders
do not arise because the microscopic coupling constants
have locally crossed into a different bulk phase. Rather,
they are induced by the unavoidable coexistence and spatial reorganization of the two primary magnetic components inside the strain-generated nematic domain wall.
This provides a particularly direct example of how bending can expose otherwise hidden composite degrees of
freedom.

\subsection{Charge-density-wave and other composite orders}
\label{sec:cdw_vestigial}

The same general mechanism applies when the primary order is a
charge-density wave rather than magnetism. A particularly relevant example
is provided by the cuprates, where charge order is observed at two
symmetry-related ordering vectors~\cite{Ghiringhelli2012,Chang2012,BlancoCanosa2013,LeTacon2014,Comin2015,Zhou2025},
\begin{equation}
    {\bf Q}_X=(Q,0,0),
    \qquad
    {\bf Q}_Y=(0,Q,0),
\end{equation}
which in a commensurate schematic description may be viewed as the
$(\pi,0)$ and $(0,\pi)$ directions. We denote the corresponding complex
CDW amplitudes by $\rho_X$ and $\rho_Y$. A state dominated by $\rho_X$
breaks rotational symmetry in a different way from one dominated by
$\rho_Y$, while the combination
\begin{equation}
    N
    =
    |\rho_X|^2-|\rho_Y|^2
    \label{eq:CDW_nematic}
\end{equation}
transforms as a nematic order parameter. The cuprate problem therefore provides another concrete example of the
central idea of this work. In a homogeneous sample, strain can be used to
select between symmetry-related stripe directions. In a bent sample, by
contrast, the sign reversal of strain forces the system to interpolate
between them within a single crystal. 

The possibility that such nematicity survives even when translational CDW
order is destroyed, e.g. by impurities, has been discussed in Refs.~\cite{Kivelson1998,Nie2014,
Nie2017}. In this case,
$N$ is a vestigial order parameter: one may have
\begin{equation}
    \langle\rho_X\rangle
    =
    \langle\rho_Y\rangle
    =
    0,
    \qquad
    \langle N\rangle
    \neq0.
\end{equation}
Orientational order can persist even when fluctuations or disorder
have destroyed long-range translational order of the underlying density
wave.

Because $N$ transforms in the same symmetry channel
as the corresponding orthorhombic strain, it couples linearly according to
\begin{equation}
    {\cal F}_{\rm strain-CDW}
    =
    -\lambda
    \int d^d r\,
    \varepsilon_{ B_{1g}}({\bf r})N({\bf r}),
    \label{eq:strain_CDW}
\end{equation}
just like in Eq.~\eqref{eq:bilinear_coupling}.
Strain therefore acts as a spatially varying anisotropy
between the two charge-ordering directions~\cite{Kim2018}. Positive and negative strain
favor the two orthogonal CDW components, respectively.

For a commensurate CDW, the composite that emerges in the neutral
plane of the bent system is 
\begin{equation}
\phi\left(z\right)=\rho_{X}\left(z\right)\rho_{Y}\left(z\right)
\end{equation}
and corresponds to a biaxial $C_{4}$-symmetric $2Q$ charge density wave,
i.e. to checkerboard order.

The situation becomes richer in the case of incommensurate CDWs where
$\rho_{X,Y}(\boldsymbol{r})$ are complex with phases $\theta_{X,Y}\left(\boldsymbol{r}\right)$:
\begin{equation}
\rho(\boldsymbol{r})=\rho_{0}+\left[\rho_{X}(\boldsymbol{r})e^{i\boldsymbol{Q}_{X}\cdot\boldsymbol{r}}+\rho_{Y}(\boldsymbol{r})e^{i\boldsymbol{Q}_{Y}\cdot\boldsymbol{r}}+\mathrm{c.c.}\right]
\end{equation}
Here, we use $\boldsymbol{Q}_{X}=\left(Q,0,0\right)$ and $\boldsymbol{Q}_{Y}=\left(0,Q,0\right)$,
related by a $C_{4}$ rotation. Nontrivial bilinears now correspond
to $H_{X,Y}=\rho_{X,Y}^{2}$ that are to second harmonics along
the $x$- and $y$-directions. In addition $B_{+}=\rho_{X}$$\rho_{Y}$
and $B_{-}=\rho_{X}^{*}\rho_{Y}$ correspond to composite smectic states with ordering vectors $\boldsymbol{Q}_{X}\pm\boldsymbol{Q}_{Y}$.
With $\rho_{X,Y}\left(z\right)=A_{X,Y}\left(z\right)e^{i\theta_{X,Y}\left(z\right)}$,
follows that near the neutral plane, $A_{X}\left(z\right)A_{Y}\left(z\right)$
behaves like $\left|\phi\left(z\right)\right|$ in Fig.~\ref{fig:phi_S}. Only disorder,
lateral edges, or a suitable in-plane surface modulation can pin the
phases.

\begin{table}[tb]
 \centering
 \small
 \setlength{\tabcolsep}{5pt}
 \renewcommand{\arraystretch}{1.25}
 \begin{tabular}{lc}
  \hline\hline
  Bilinear & Physical channel \\
  \hline
  $R=\Psi^\dagger \tau_0 \Psi$ & pairing amplitude \\
  $\mathbf{N} = \left(\Psi^\dagger \tau_z \Psi, -\Psi^\dagger \tau_x \Psi \right)$ & Nematic \\
  $\chi=\Psi^\dagger \tau_y \Psi$
      & TRSB, axial \\
  \hline\hline
 \end{tabular}
  \caption{Bilinears formed from the two components $\psi_1$ and $\psi_2$ of the superconducting order parameter $\Psi=(\psi_1,\psi_2)^T$, where the $\tau_\alpha$ are the Pauli matrices with $2\times 2$ unit matrix for $\alpha=0$. $R$ stands  for the symmetry-preserving pairing amplitude square. Nonzero $\mathbf{N}$ corresponds to nematic superconductivity, where for a tetragonal system the two components of $\mathbf{N}$ transform under the $B_{1g}$ and $B_{2g}$ irreducible representation of $D_{4h}$, respectively. For a trigonal system $\mathbf{N}$ transforms under the two-dimensional irrep $E_g$ of $D_{3d}$. $\chi$ is odd under time-reversal and even under inversion, i.e. it behaves like the component of an axial vector.
   }
  \label{tab:sc_bilinears}
\end{table}

\begin{figure}[tb]
    \centering
    \includegraphics[width=\linewidth]{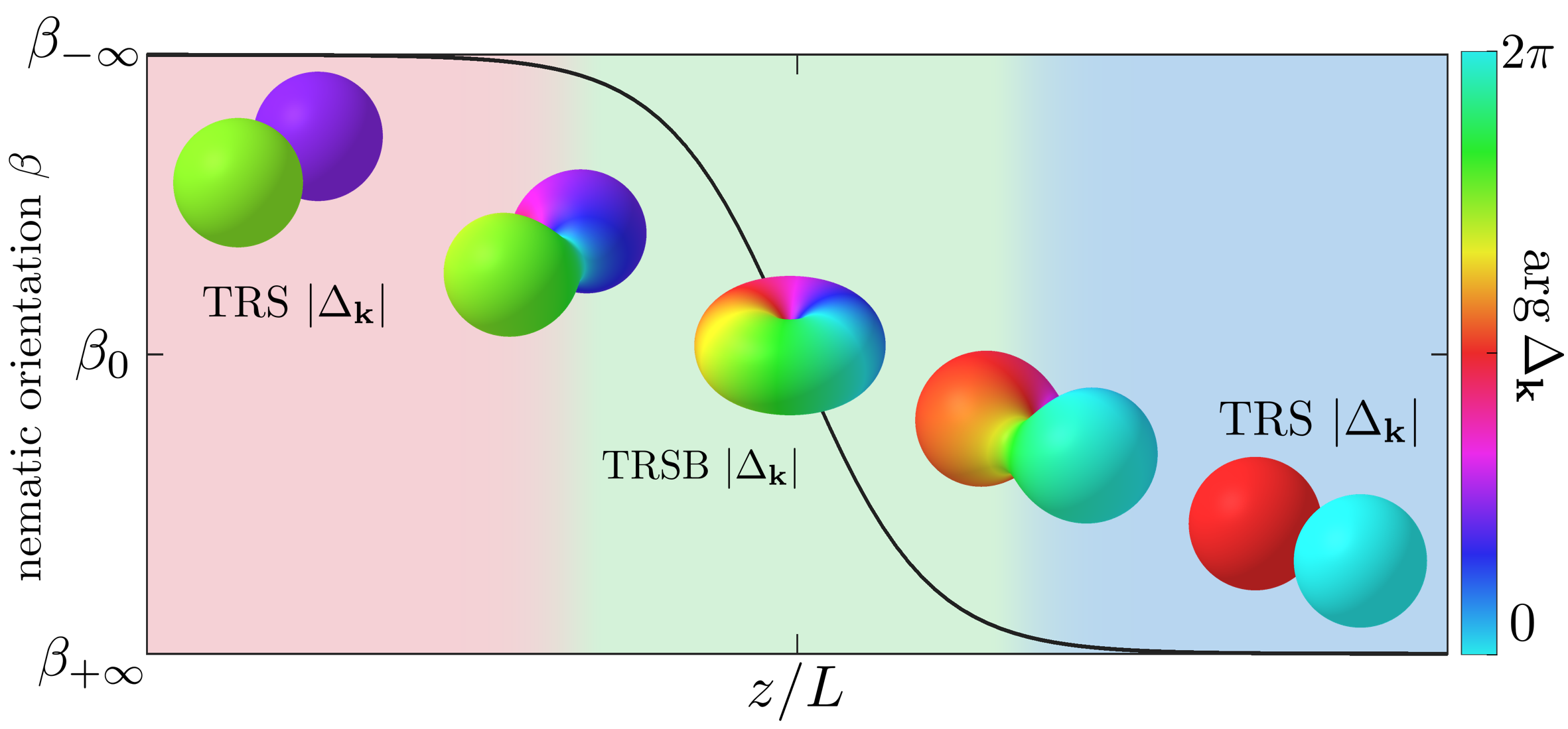}
    \caption{Schematic of interfacial axial order between strain-selected nematic superconducting domains. Sign-changing strain selects distinct TRS nematic orientations on opposite sides of the neutral plane, while the interface supports TRSB axial order. The black curve illustrates the spatial variation of the nematic orientation $\beta$. Insets show schematic three-dimensional orbital paring-gap profiles for both TRS and TRSB states. The surface radius represents $|\Delta_{\mathbf{k}}|$, and color indicates $\arg\Delta_{\mathbf{k}}$ modulo $2\pi$.}
    \label{fig:SC_schematics}
\end{figure}

\subsection{Time-reversal symmetry breaking in nematic superconductors}
\label{sec:TRSBSC}

The preceding examples show how a sign-changing strain field can reveal the internal structure of an electronic order in the particle-hole sector. Strain favors different nematic domain orientations on opposite sides of the neutral plane, bending forces them to coexist and reorganize within an interface, where additional new order  emerges.  Multi-component superconductors provide another direct setting in which to examine this mechanism: strain couples to a nematic bilinear of the
Cooper-pair field. We find that  the resulting nematic domain wall can lead to a time-reversal symmetry breaking (TRSB) superconducting order at the interface. This provides a controlled realization of TRSB discussed in the context of twin boundaries and strain defects~\cite{Arahata2013,Watashige2015,Willa2021,Yuan2021}.  

We  follow the superconducting texture within the phase-coherent superconducting phase, described by the two-component order parameter $\Psi=(\psi_1,\psi_2)^T$, where the components are the complex amplitudes of two symmetry-related pairing basis functions. We write 
\begin{equation}
    \Psi = \rho e^{i\varphi} \left[ \cos \frac{\eta}{2}\begin{pmatrix}
        \cos \frac{\beta}{2}\\
        -\sin \frac{\beta}{2}
    \end{pmatrix} + i  \sin \frac{\eta}{2}\begin{pmatrix}
        \sin \frac{\beta}{2}\\
        \cos \frac{\beta}{2}
    \end{pmatrix} \right],
    \label{eq:sc_order_para}
\end{equation}
where $\rho>0,-\pi/2\leq \eta \leq\pi/2, 0\leq \beta,\theta\leq 2 \pi$ is assumed.
Here $\rho$ is the superconducting amplitude and $\varphi$ is the common phase. The angles $\beta$ and $\eta$ describe the nematic orientation and the departure from a time-reversal-symmetric(TRS) state, respectively. These follow from the gauge-invariant bilinears of Table~\ref{tab:sc_bilinears}
\begin{eqnarray}
    \mathbf{N} & =& \rho^2 \cos \eta (\cos \beta,\sin \beta) \nonumber \\
    \chi &=& \rho^2 \sin \eta.
    \label{eq:scbilinears}
\end{eqnarray}
The nematic bilinear $\mathbf{N}$ is invariant under time reversal, whereas  $\chi$ changes sign.  For $\eta = 0$, the superconducting order parameter is real up to a common phase and preserves time-reversal symmetry. A non-zero $\chi$ identifies a TRSB state. We refer to $\chi$ as a time-reversal-odd axial order parameter. We use the term axial instead of chiral to emphasize that the time-reversal-breaking bilinear $\chi=\Psi^\dagger\tau_y\Psi$ transforms as the $z$- component of an axial vector, being even under inversion and odd under time reversal, rather than as the inversion-odd, time-reversal-even pseudo-scalar associated with structural chirality.

Strain field directly couples to the $\mathbf{N}$ bilinear, favoring different nematic orientations on opposite sides of the neutral plane.
As shown in Fig.~\ref{fig:SC_schematics}, the superconducting order parameter must interpolate between these domains, allowing both its amplitude and its orientation to vary. We show below that a TRSB interface can be energetically favorable in both tetragonal and trigonal nematic superconductors whose bulk homogeneous states are TRS. The trigonal case is of particular experimental relevance given the evidence for nematic superconductivity in the doped trigonal Bi$_2$Se$_3$~\cite{Matano2016,Yonezawa2017,Pan2016,Du2017,Asaba2017,Willa2018,Tao2018,Kostylev2020,Cho2020}.

\subsubsection{Tetragonal nematic superconductors}
We consider a tetragonal crystal with point group $D_{4h}$ and a two-component superconducting order parameter $\Psi$ transforming according to the $E_g$ or $E_u$ irreducible representation. Representative pairing basis functions are $(d_{xz},d_{yz})$ or $(p_x,p_y)$, respectively.
For either pairing representation, the superconducting bilinears decompose into 
\begin{equation}
    \label{eq:tetra_decomp}
    E^*_{u,g}\otimes E_{u,g} =  A_{1g}\oplus A_{2g}\oplus B_{1g}\oplus B_{2g}.
\end{equation}
 In particular, the components $N_1$ and $N_2$ of the vector $\mathbf{N}$ in Eq.~\eqref{eq:scbilinears} transform as the distinct representations $B_{1g}$ and $B_{2g}$, respectively. We focus on $B_{1g}$ nematic superconductivity, whose two degenerate, time-reversal-symmetric bulk states are $\Psi\propto(1,0)^T$ and $\Psi\propto(0,1)^T$. The corresponding problem for $B_{2g}$ nematic states, $\Psi\propto(1,\pm1)^T$, coupled to $B_{2g}$ strain can be mapped onto this case by rotating the order-parameter basis by $\pi/4$ and transforming the free-energy coefficients accordingly. We therefore present the analysis explicitly for the $B_{1g}$ case only.

To determine the resulting superconducting texture and whether it breaks time-reversal symmetry, we consider the Ginzburg–Landau free-energy density
\begin{equation}
\label{eq:tetragonal_free_energy}
\begin{aligned}
f &=
\frac{d}{2}\left(
|\partial_z\psi_1|^2 + |\partial_z\psi_2|^2
\right)
+\frac{r}{2}\left(
|\psi_1|^2 + |\psi_2|^2
\right)\\
&+\frac{u}{4}\left(
|\psi_1|^2 + |\psi_2|^2
\right)^2
+\frac{u'}{4}
\left(\psi_1^*\psi_2-\psi_2^*\psi_1\right)^2\\
&+\frac{u''}{4}|\psi_1|^2|\psi_2|^2
-\lambda \varepsilon(z) \left(
|\psi_1|^2-|\psi_2|^2
\right),
\end{aligned}
\end{equation}
Here $d>0$ is the gradient stiffness, $r<0$ places the unstrained system in the superconducting phase, and $\varepsilon(z)$ is the imposed $B_{1g}$ strain. The temperature dependence can be captured by writing $r=a(T-T_c)/T_c$ with $T_c$ being the superconducting temperature. We focus on the parameter regime $u>0$ and $u''>\max(4u',0)$, which ensures quartic stability and favors two degenerate, time-reversal-symmetric bulk states, $\Psi\propto(1,0)^T$ and $\Psi\propto(0,1)^T$. These states correspond to the two orientations of a $B_{1g}$ nematic superconductor.

\begin{figure}[tbp]
    \centering
    \includegraphics[width=\linewidth]{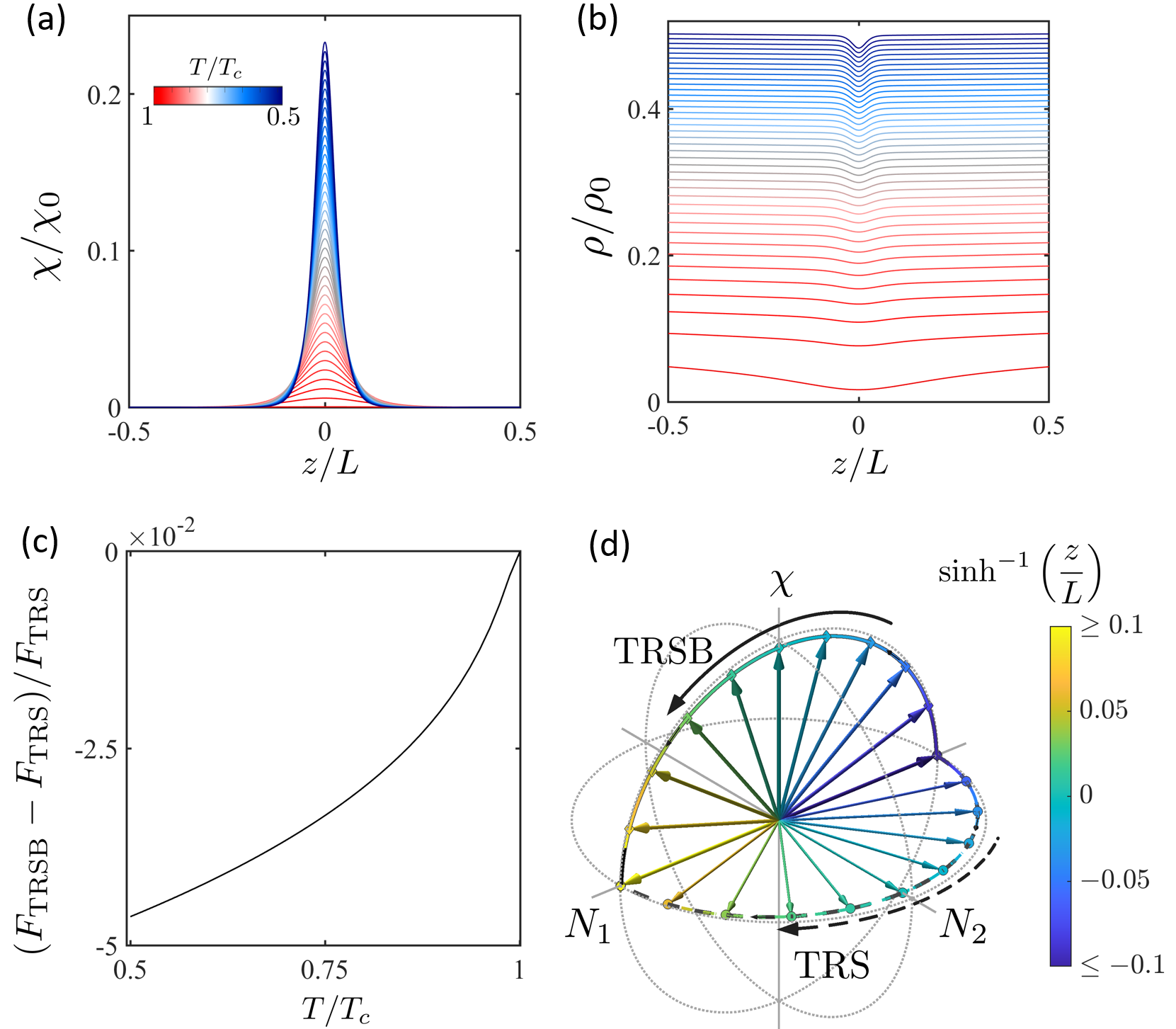}
    \caption{ Time-reversal-breaking axial order at an interface in a tetragonal nematic superconductor. (a-b) Spatial profiles of the axial order parameter $\chi$ in unit of $\chi_0 = 2a/u$ and superconducting amplitude $\rho$ in unit of $\rho_0 = \sqrt{2a/u}$ for the lower-energy TRSB branch, shown at   temperatures between $T/T_c=0.5$ and $1$. Here $\chi_0=\rho_0^2$. Axial order develops near the neutral plane and vanishes in the nematic bulk domains. (c) Relative free-energy difference $(F_{\mathrm{TRSB}}-F_{\mathrm{TRS}})/F_{\mathrm{TRS}}$, where $F$ denotes the bulk-subtracted excess free energy defined in the Appendix~\ref{app:tetragonal_gl}. The negative values indicate that the TRSB branch is energetically favored. (d) Bloch-space trajectories of the TRS (dashed line,) and TRSB (solid line) solutions at $T/T_c=0.5$. Each radial vector has length $\rho/\rho_0$, azimuth $\beta$, and elevation $\eta$. The TRS trajectory remains in the nematic equatorial plane, whereas the TRSB trajectory acquires a finite axial component. Colors encode position through $\mathrm{arcsinh}(z/L)$, with saturation beyond the indicated color limits. Parameters are given in Appendix~\ref{app:tetragonal_gl}.}
    \label{fig:tetra_SC}
\end{figure}

Taking $\lambda>0$, positive strain favors the $\psi_1$ component, whereas negative strain favors $\psi_2$. A strain profile that changes sign across the neutral plane therefore selects different nematic domains on the two sides and creates an interface between them.  Using our parameterization Eq.~\eqref{eq:sc_order_para}, minimization with respect to $\varphi$ gives 
\begin{equation}
\label{eq:global_phase_eq}
\partial_z\varphi+\frac12\sin\eta\,\partial_z\beta
 =0,
\end{equation}
in the absence of an imposed transport current along $z$. 

Substituting this relation into the free-energy density gives the reduced free energy as function of $(\rho,\beta,\eta)$,
\begin{equation}
\label{eq:sc_f_red}
\begin{aligned}
f={}&
\frac d2(\partial_z\rho)^2
+\frac{\rho^2 d}{8}
\left[
(\partial_z\eta)^2+
\cos^2\eta\,(\partial_z\beta)^2
\right]
+\frac r2\rho^2+\frac u4\rho^4
\\
&+\frac{\rho^4}{16}
\left[
u''\cos^2\eta\,\sin^2\beta
+(u''-4u')\sin^2\eta
\right]\\
&-\lambda\varepsilon(z)\rho^2\cos\eta\cos\beta.
\end{aligned}
\end{equation}

To determine the superconducting texture connecting the two strain-selected nematic domains, we solve the Euler–Lagrange equations derived from Eq.~\eqref{eq:sc_f_red} numerically. We focus on $0<u'<u''/4$ and compare TRS and TRSB solutions subject to the same bulk amplitudes and nematic orientations, with no externally imposed phase difference between the domains. The TRS branch is obtained by constraining $\eta=0$, while the TRSB branch is found by allowing all three fields, $\rho$, $\beta$, and $\eta$, to vary, starting from a localized nonzero $\chi$. Implementation details are given in Appendices\ref{app:tetragonal_gl} and \ref{app:trigonal_gl}. To establish the energetic preference between the two branches, Fig.~\ref{fig:tetra_SC}(c) compares their relative excess free energies, $(F_{\mathrm{TRSB}}-F_{\mathrm{TRS}})/F_{\mathrm{TRS}}$.  The TRSB branch has lower energy throughout the temperature range shown.  

Fig.~\ref{fig:tetra_SC}(a-b) show the axial order parameter $\chi$ and superconducting amplitude $\rho$ profiles of the lower-energy TRSB branch at different temperatures. A finite TRSB phase develops near the neutral plane and decays toward the time-reversal-symmetric nematic domains on either side. Its peak increases upon cooling. The superconducting amplitude is reduced near the interface but remains finite, so axial order develops within a continuous superconducting texture. Time reversal changes the sign of $\chi$ while leaving the bulk nematic states unchanged, producing two degenerate interfacial states; one partner is shown in Fig.~\ref{fig:tetra_SC}(a). The distinction between the two textures is illustrated geometrically in Fig.~\ref{fig:tetra_SC}(d). The TRS solution connects the strain-selected bulk states within the $N_1$–$N_2$ plane along the equator, whereas the TRSB solution follows a path with a finite axial component. Both trajectories connect the same nematic endpoints. 

The origin of the energy reduction along the TRSB path can be understood by varying the relative phase between the two superconducting components. For a TRS configuration $\Psi_\mathrm{TRS} =(a(z),b(z))^T$, where $a(z)$ and $b(z)$ are real, we construct the trial configuration $\Psi_\mathrm{TRSB} = (a ,e^{i\delta} b)^T$ with a constant phase $\delta \neq 0,\pi$. Such a configuration is compatible with the boundary condition $\eta(\pm \infty) = 0$ as long as either $a$ or $b$ vanishes at the end points. The free-energy difference of these two configurations is 
\begin{equation}
    \label{eq:DelF_tetra}
    F[\Psi_\mathrm{TRSB}] - F[\Psi_\mathrm{TRS}]=-u'\sin^2\delta \int dz\, (a b)^2.
\end{equation}
For $u'>0$, the energy is minimized within this family at $\delta=\pm\pi/2$ and is strictly lower than the TRS energy whenever the two components $a$ and $b$ overlap. 
This produces interfacial axial order and breaks time-reversal symmetry.
For $u'<0$ the transition between the $(1,0)$ and $(0,1)$ nematic states is through a rotation of the nematic axis reaching $(1,\pm 1)$ at $z=0$.

At low temperatures, the modest amplitude variation in Fig.~\ref{fig:tetra_SC}(b) motivates the fixed-amplitude approximation $\rho(z)\simeq\bar\rho$, where $\bar\rho$ denotes the characteristic superconducting amplitude at the temperature considered. In the fixed-amplitude, weak strain
approximation,   Appendix~\ref{app:tetragonal_gl} gives the axial profile,,
\begin{equation}
    \label{eq:tetra_chi_profile}
   \chi = \pm \frac{\bar{\rho}^2}{\cosh(z/\ell_\chi)},\qquad 
   \ell_\chi = \sqrt{\frac{d}{2\bar{\rho}^2 (u''/4-u')}}.
\end{equation}
 
\subsubsection{Trigonal nematic superconductors}
\label{sec:trigonal}
\begin{figure}[tbp]
    \centering
    \includegraphics[width=\linewidth]{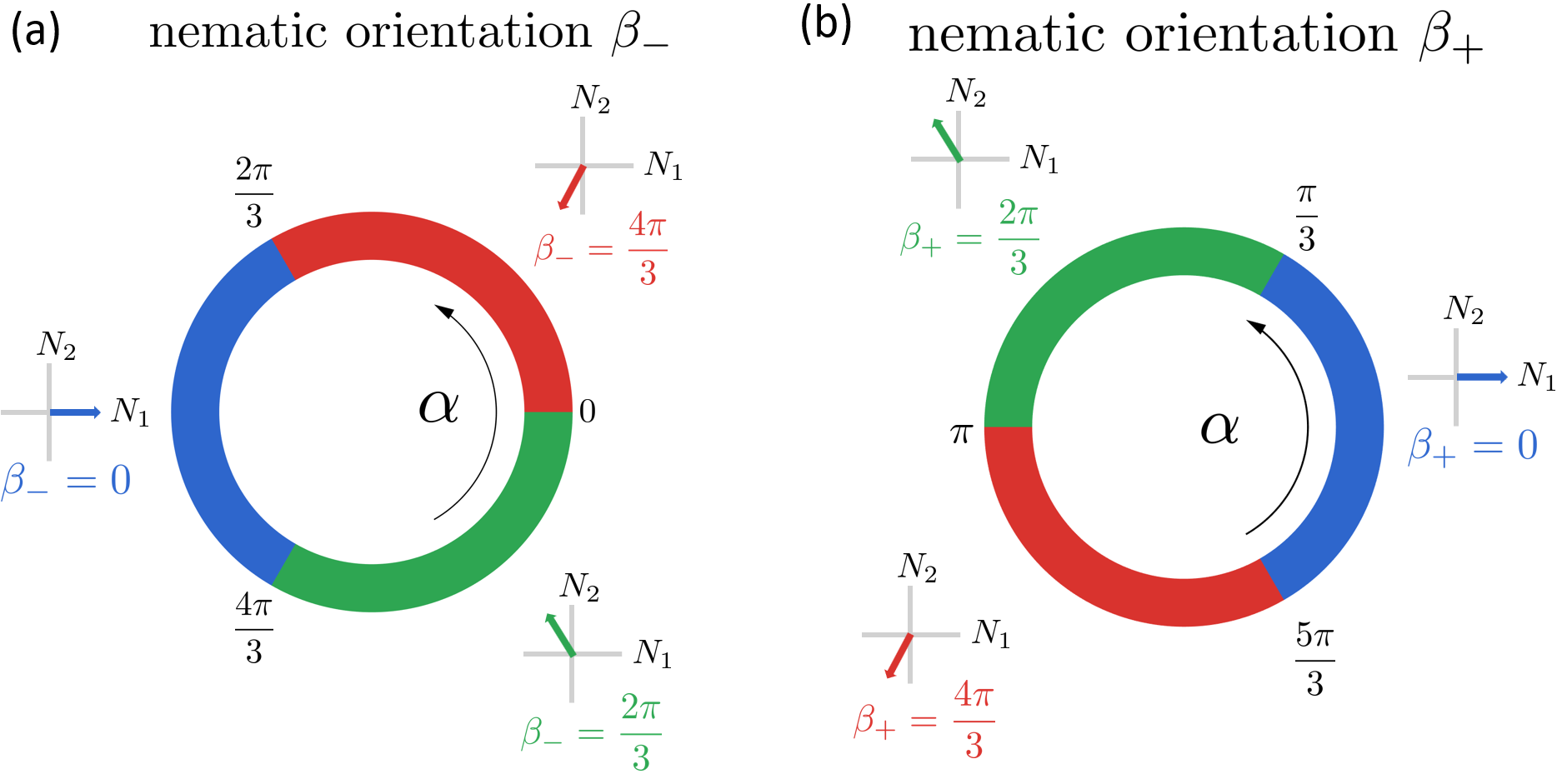}
    \caption{Strain-selected bulk nematic orientations in a trigonal superconductor. Equilibrium nematic orientation (a) $\beta_-$ and (b) $\beta_+$ at $z\to\mp \infty$, as functions of  strain angle $\alpha$ with $2h_\infty/a = 10^{-4}$.   In the weak-strain regime shown, crystalline anisotropy pins the orientations close to $0$, $2\pi/3$, and $4\pi/3$, while varying $\alpha$ switches the selected nematic domain. Parameters used here are $T = 0.9T_c, g/u=0.01,4av/3u^2=0.1$. }
    \label{fig:tri_bulk}
\end{figure}

We next consider a trigonal crystal with point group $D_{3d}$ and a two-component superconducting order parameter $\Psi$ transforming according to the $E_g$ or $E_u$ irreducible representation. An important example is electron-doped $\mathrm{Bi_2Se_3}$, including $\mathrm{Cu}_x \mathrm{Bi_2Se_3}$ and related $\mathrm{Sr}$- and $\mathrm{Nb}$-doped compounds which are low-carrier density superconductors with $T_c$ of order $3$--$4\mathrm{K}$~\cite{Matano2016,Yonezawa2017,Pan2016,Du2017,Asaba2017,Willa2018,Tao2018,Kostylev2020,Cho2020,Fu2014,Venderbos2016,Hecker2018}. We use
this system as a concrete example of how inhomogeneous
strain can generate novel states of order in an experimentally accessible system.
Indeed,  experimental evidence for control of nematic superconducting domains by uniaxial strain in Sr$_x$Bi$_2$Se$_3$ was reported in Ref.~\cite{Kostylev2020}.

The gauge-invariant superconducting bilinears of Table~\ref{tab:sc_bilinears} decompose according to
\begin{equation}
    \label{eq:tri_decomp}
    E^*_{u,g}\otimes E_{u,g} = A_{1g}\oplus A_{2g}\oplus E_g.
\end{equation}
Here, $(N_1,N_2)$ form an $E_g$ doublet. Consequently, their quadratic invariant $N_1^2+N_2^2$ does not distinguish nematic orientations, and the leading crystalline pinning appears at sixth order in $\Psi$, selecting three symmetry-related orientations. A sign-changing strain field can favor distinct nematic domains on opposite sides of the neutral plane. We investigate whether the superconducting texture connecting these domains develops interfacial axial order.

Using the bilinears introduced above, we consider the  Ginzburg–Landau free-energy density~\cite{Fu2014,Venderbos2016}
\begin{equation}
    \label{eq:GL_tri}
    \begin{aligned}
        f =& \frac{d}{2}(|\partial_z \psi_1|^2 + |\partial_z \psi_2|^2) + \frac{r}{2}\rho^2 + \frac{u}{4} \rho^4 \\
        +&\frac{g}{4} \chi^2 + \frac{v}{6}[\rho^6 - \mathrm{Re}(N_1+i N_2)^3] - \mathbf{h}(z)\cdot \mathbf{N}
    \end{aligned}
\end{equation}
where $d,u,g,v>0$, $r = a(T-T_c)/T_c<0$. The positive coefficient $g$ penalizes TRSB. The strain field couples to the nematic bilinear through an effective field $\mathbf h(z)=(h_1,h_2)$, which incorporates the strain-coupling constants. For $D_{3d}$ symmetry, the strain tensor contains two $E_g$ doublets, $\boldsymbol\epsilon_1=(\epsilon_{xx}-\epsilon_{yy},-2\epsilon_{xy})$ and $\boldsymbol\epsilon_2=(2\epsilon_{yz},-2\epsilon_{xz})$, describing in-plane anisotropic strain and shear involving the trigonal axis, respectively. We have performed finite-element simulations using the elastic constants of Bi$_2$Se$_3$ reported in Ref.~\cite{Gao2016} and find that bending induces $\epsilon_{yz}\sim 0.3- 0.4 (\epsilon_{xx}-\epsilon_{yy})$,
see Appendix~\ref{app:fem}. The components of the strain doublet  couple linearly to the nematic doublet $\mathbf N$, giving $\mathbf{ h}(z)=-\kappa_1\boldsymbol\epsilon_1(z)-\kappa_2\boldsymbol\epsilon_2(z)$, where $\kappa_1$ and $\kappa_2$ are phenomenological coupling constants. Our conclusions do not depend sensitively on the precise values of these parameters. Indeed, most of our analytical arguments assume infinitesimal strain, which merely lifts the degeneracy of the order parameter in the upper and lower parts of the sample.
In the following discussion, we assume a simple form $\mathbf{h}(z) = h_\infty\tanh(z/L)(\cos \alpha,\sin \alpha)$. The sixth-order term is nonnegative, and both angular contributions attain their minimum at
\begin{equation}
    \label{eq:tri_degenerate}
    \begin{aligned}
         \eta = 0,\qquad \beta = 0,\frac{2\pi}{3},\frac{4\pi}{3}.
    \end{aligned}
\end{equation}
For $r<0$, these angles describe the three degenerate, TRS nematic ground states of the unstrained homogeneous system. Strain generically lifts this degeneracy through the coupling $-\mathbf h\cdot\mathbf N$. Fig.~\ref{fig:tri_bulk} shows the resulting bulk orientations $\beta_\pm$ selected by the opposite asymptotic fields $\mathbf h_\pm=\pm h_\infty(\cos\alpha,\sin\alpha)$. In the weak-strain regime shown, the orientations remain close to the crystalline values in Eq.\eqref{eq:tri_degenerate}, while varying $\alpha$ switches the selected nematic domains. For general $\alpha$, the nematic ground state is unique while for special angles $\alpha=n\pi/3$, with integer $n$, two neighboring nematic orientations remain degenerate at one endpoint.  

To determine this texture, we consider spatial variations along the trigonal axis. Minimization over the common phase again gives Eq.~\eqref{eq:global_phase_eq}. Using the parameterization in Eq.~\eqref{eq:sc_order_para}, the reduced free-energy density becomes
\begin{equation}
\label{eq:tri_GL_f}
\begin{aligned}
f={}&
\frac d2(\partial_z\rho)^2
+\frac{d\rho^2}{8}
\left[
(\partial_z\eta)^2
+\cos^2\eta (\partial_z\beta)^2
\right]\\
&+\frac r2\rho^2+\frac u4\rho^4
+\frac g4\rho^4\sin^2\eta\\
&+\frac v6\rho^6
\left(1-\cos^3\eta\cos3\beta\right)
-\rho^2\cos\eta h_\parallel(z,\beta),
\end{aligned}
\end{equation}
where $h_\parallel(z,\beta)=h_\infty \tanh(z/L)\cos(\beta-\alpha)$ is the projection of the strain field onto the nematic direction.

\begin{figure}[tbp]
    \centering
    \includegraphics[width=\linewidth]{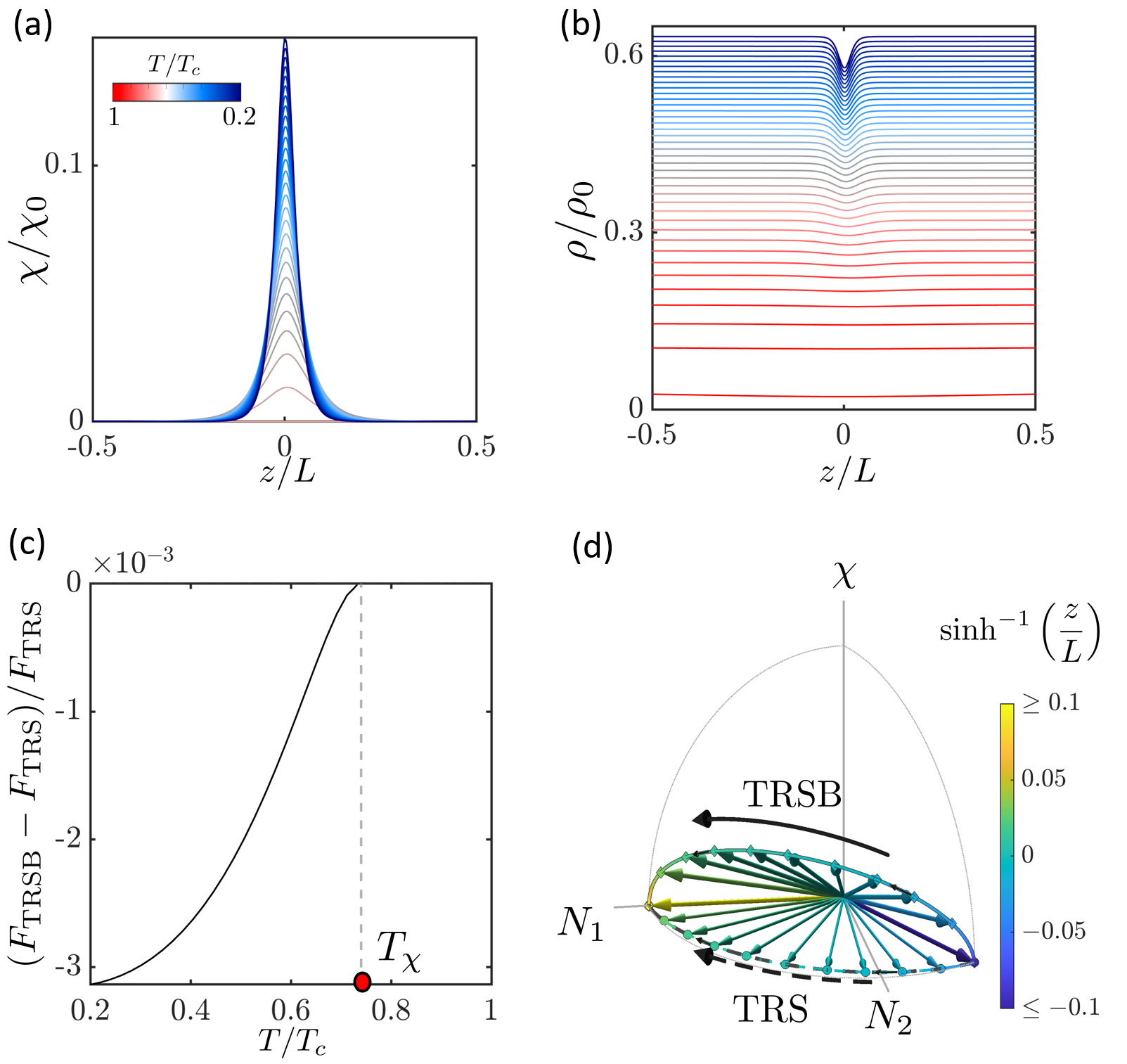}
    \caption{ Interfacial axial order in a trigonal nematic superconductor. (a–b) Spatial profiles of the axial order parameter $\chi$ in units of $\chi_0=2a/u$ and superconducting amplitude $\rho$ in units of $\rho_0=\sqrt{2a/u}$ for the lower-energy solution, shown at temperatures between $T/T_c=0.2$ and $1$. Here $\chi_0=\rho_0^2$. Below the interfacial transition temperature, axial order develops near the neutral plane and vanishes in the nematic bulk domains. (c) Relative free-energy difference $(F_{\mathrm{TRSB}}-F_{\mathrm{TRS}})/F_{\mathrm{TRS}}$. Negative values indicate that the TRSB branch is energetically favored. The gray dashed line marks the axial instability of the TRS branch. (d) Bloch-space trajectories of the TRS (dashed line) and TRSB (solid line) solutions at $T/T_c=0.2$. Parameters are given in  Appendix~\ref{app:trigonal_gl}}
    \label{fig:tri_SC}
\end{figure}

We solve the Euler–Lagrange equations derived from Eq.~\eqref{eq:tri_GL_f} subject to the strain-selected bulk boundary conditions. As in the tetragonal case, we obtain the TRS branch by imposing $\eta=0$ and search for a TRSB branch by allowing $\rho$, $\beta$, and $\eta$ to vary, starting from a localized seed with nonzero $\chi$. Both branches connect the same bulk nematic states, with no externally imposed superconducting phase difference. The numerical results reveal a separate interfacial transition below the superconducting transition temperature. We denote its critical temperature by $T_\chi$, where the lowest localized axial fluctuation of the TRS wall becomes unstable. As shown in Fig.~\ref{fig:tri_SC}(c), $T_\chi\simeq0.74T_c$, indicated by the gray dashed line. Below this temperature, the TRSB branch has lower total free energy. The separate phase transition to a TRSB states below $T_c$ for the trigonal geometry may facilitate the identification of the new gradient order.

Fig.~\ref{fig:tri_SC}(a–b) show how this instability changes the superconducting texture. Below $T_\chi$, axial order develops near the neutral plane and grows upon cooling, while vanishing in the nematic bulk domains. The superconducting amplitude is reduced near the interface but remains finite. The transition therefore introduces axial order within an already superconducting texture. In the Bloch representation of Fig.~\ref{fig:tri_SC}(d), the TRS trajectory remains in the $N_1$–$N_2$ plane, whereas the TRSB trajectory tilts out of the plane and develops a finite $\chi$ component while connecting the same endpoints.   

The finite threshold $T_\chi$ reflects a competition between the positive quartic energy cost of axial order and the energy required to rotate between crystalline nematic orientations.  The positive $g$ term penalizes a nonzero $\chi$, and spatial variations of $\eta$ introduce an additional gradient cost. However, leaving the nematic plane can reduce both the gradient energy associated with the rotation of $\beta$ and the crystalline anisotropy barrier encountered along the TRS path.

This competition can be understood by asking whether the TRS domain wall can lower its energy by developing a small axial order parameter. The TRS solution is obtained by minimizing the free energy with $\eta=0$, allowing the amplitude and nematic orientation to vary across the interface. To test its stability when this restriction is lifted, we introduce a small departure $\eta(z)=\epsilon w(z)$, where $\epsilon\ll1$ and $w(z)$ vanishes in the bulk. This distortion produces local axial order, $\chi(z)\simeq\rho^2(z)\epsilon w(z)$, while preserving the same time-reversal-symmetric states at both ends. If it lowers the free energy, the TRS wall is unstable toward a texture with axial order.

An analytical estimate follows by taking a constant amplitude $\rho=\bar\rho$ and neglecting the weak strain within the wall, while retaining the strain-selected bulk orientations as boundary conditions. For a wall connecting neighboring nematic minima, the most favorable distortion of fixed norm in the axial channel gives
\begin{equation}
    F-F_{\mathrm{TRS}}
=
\epsilon^2\frac{\bar\rho^4}{12}
\left(3g-v\bar\rho^2\right)
\int dz\,w^2(z),
\end{equation}
as derived in Appendix~\ref{app:trigonal_gl}. The TRS wall therefore becomes unstable when $v\bar\rho^2>3g$.

Using the weak-strain bulk amplitude $\bar\rho^2\simeq-r/u$ and $r=a(T/T_c-1)$ gives
\begin{equation}
    T_\chi\simeq T_c\left(1-\frac{3ug}{av}\right),
\end{equation}
provided the resulting temperature is positive. For the numerical parameters used in Fig.~\ref{fig:tri_SC}, this estimate yields $T_\chi\simeq0.80T_c$, which is in reasonable agreement with  the numerical value $T_\chi\simeq0.74T_c$.  

Near a continuous onset, the distortion that first becomes unstable determines the leading axial order parameter profile,
\begin{equation}
    \chi(z)\simeq
\pm \chi_m(T)\,
\operatorname{sech}^{2/3}\!\left(\frac{z}{\ell_\chi}\right),
\quad
\ell_\chi=\sqrt{\frac{d}{6v\bar\rho^4}}.
\end{equation}
Here the amplitude $\chi_m(T)$ vanishes at $T_\chi$ and is determined by nonlinear terms below the transition. The two signs describe time-reversal partners. 

\subsection{Strain gradients and vestigial order}
Thus far, we have focused on the emergence of additional symmetry breaking in the neutral plane of systems with long-range magnetic or superconducting order. The essential ingredient is a two-component order parameter describing the homogeneous bulk, whose degeneracy can be lifted by strain. Such systems may also support vestigial order: the primary order parameters exhibit no long-range order, $\left\langle \mathbf{S}_{X}\right\rangle =\left\langle \mathbf{S}_{Y}\right\rangle =0$
or $\left\langle \Psi\right\rangle =0$, while the composite order parameters listed in Tables~\ref{tab:magnetic_bilinears}  and \ref{tab:sc_bilinears} acquire finite expectation values. As discussed in detail in Refs.~\cite{Fernandes2010,Fernandes2016,Fernandes2019}, vestigial order typically occurs in a relatively narrow temperature window above the onset of primary order. It can be understood as a partial melting of the primary ordered state, in which composite order survives to slightly higher temperatures.

Much of our reasoning extends naturally to this regime. In the vestigial nematic phase associated with stripe magnetism, we would expect double-$Q$ order to emerge in the neutral plane. For tetragonal superconductors, we would similarly expect time-reversal symmetry breaking in the neutral plane, even in the absence of superconductivity. A particularly intriguing possibility arises in nematic superconductors on lattices with sixfold rotational symmetry, where vestigial charge-$4e$ superconductivity emerges as a closely competing phase~\cite{FernandesFu2021}. The possibility of charge-$4e$ order near nematic domains, discussed in Ref.~\cite{Hecker2024}, suggests that controlled bending could provide a route to stabilizing this phase in the neutral plane.

\section{Discussion and Conclusions}
\label{sec:conclusions}

Controlled bending provides a route to creating electronic interfaces within a single crystal and using them to stabilize order absent from the surrounding bulk. A sign-changing strain field selects distinct symmetry-related configurations on opposite sides of the sample, while finite order-parameter stiffness connects them through an interface. Their competition can activate additional internal degrees of freedom near the neutral plane, where the relevant strain-induced conjugate field vanishes but the neighboring domains continue to impose competing boundary conditions. This combination establishes spatially varying strain as a means of both creating and probing collective electronic states. At the same time we emphasize that hidden and unintentional strain fields can exist in real samples around defects such as cracks, dislocations and intergrowths. Such strain gradients imprinted by a materials microstructure into the electron fluid may induce gradient effects accidentally and should be considered when contradictory results challenge a rigorous understanding of the microscopics - for example the presence or absence of TRSB in superconductors. Within the gradient matter framework developed here, controlled bending allows us to vary the magnitude and sign of the imposed strain gradient and thereby probe its role in electronic symmetry breaking systematically.

Our examples demonstrate how this principle operates across different classes of quantum materials. In stripe magnets, the overlap of the two magnetic components produces a localised double-$Q$ texture with checkerboard charge order or spin vorticity, depending on their relative spin orientation. Analytical profiles and numerical solutions relate the amplitude and width of this order to the underlying stiffness and competing-order anisotropy. Piezomagnetic coupling provides a complementary application: an external magnetic field enables strain to select and pin altermagnetic domains in symmetry-allowed channels.
Nematic superconductors illustrate how an interface can acquire an additional broken symmetry while both neighboring domains remain time-reversal symmetric. In the tetragonal regime studied here, a nonlinear coupling favoring a complex relative phase stabilizes axial, time-reversal symmetry breaking order in the region where the two superconducting components overlap. In trigonal crystals, the reduction in orientational and gradient energy can overcome the positive local cost of axial order, producing a separate interfacial transition below the superconducting transition.  Analytical  criteria and profiles for gradient order are complemented and supported by numerical solutions of the non-linear Ginzburg-Landau equations and identify the energetic competition governing gradient order. Together, these results show how the internal structure of a multi-component order parameter determines the states accessible at a controlled interface.

The experimental setting  we propose is concrete and feasible. Bent crystalline microcantilevers already provide access to controlled, sign-changing strain profiles~\cite{Diaz2022}, and our finite-element calculations establish the corresponding strain geometry. The demonstrated control of nematic superconducting domains by uniaxial strain in Sr$_x$Bi$_2$Se$_3$~\cite{Kostylev2020} supplies a further ingredient for implementing the proposed superconducting interfaces, striped magnetic order with nearby $C_4$-symmetric $2Q$-order in the phase diagram has been identified in Ba$_{1-x}$K$_x$Fe$_2$As$_2$~\cite{Bohmer2015}, strongly suggesting that our approach allows to realize both states in a single sample via homogeneous bending. Finally,  our weak-strain analysis shows that, once distinct bulk domains are selected, the interfacial state can be governed primarily by the intrinsic stiffness and anisotropies of the electronic order. Hence, its formation  does not rely on a large local distortion at the neutral plane.

{
Experimental probes of gradient order must access the ordered layer near the strain-neutral plane and distinguish its response from the surrounding bulk. The exploration of surface phenomena demonstrates that two-dimensional electronic states can be spectroscopically resolved despite the much larger number of bulk degrees of freedom. Surface superconductivity~\cite{Zhang2010SurfaceSC}, surface magnetic order~\cite{Chikina2014SurfaceMagnetism}, and topological surface states ~\cite{Roushan2009SurfaceStates,Hsieh2009SpinTexture} provide familiar examples of such low-dimensional physics. At exposed surfaces, scanning tunneling microscopy and angle-resolved photoemission spectroscopy selectively probe electronic states, providing complementary real- and momentum-space information. 

The additional experimental challenge for gradient order is that the ordered layer is buried within a single crystal. Studies of oxide heterostructures provide useful methodological precedents for probing buried electronic states. 
In oxide heterostructures, X-ray absorption spectroscopy~\cite{Salluzzo2009}, optical second-harmonic generation~\cite{Savoia2009}, torque magnetometry~\cite{Li2011}, and superconducting transport~\cite{Biscaras2010} have revealed electronic reconstruction and collective order beneath the surface. Studies of metallic magnetic interfaces likewise demonstrate the value of spin-sensitive transport and magnetic resonance for investigating interfacial phenomena~\cite{Hellman2017}. These approaches offer complementary starting points for gradient quantum matter, with the choice of measurement guided by the symmetry of the expected order and its contrast with the surrounding bulk.

For the magnetic examples, measurements sensitive to the additional ordering wave-vector or spin-vorticity symmetry offer routes to identifying the double-$Q$ contribution within the stripe-ordered sample. For nematic superconductors, the central target is interfacial time-reversal symmetry breaking within an already superconducting state. The separate transition predicted for trigonal crystals provides a particularly useful temperature scale for seeking a spontaneous time-reversal-odd response. Collective-mode spectroscopy could also provide a complementary probe of multicomponent TRSB superconductivity through resonances associated with relative-amplitude and relative-phase fluctuations of the order parameter~\cite{Poniatowski2022}.
Comparing measurements under different bending configurations can further test the connection between an observed signal and the imposed domain structure. The calculated spatial profiles provide a basis for estimating the interfacial contribution and selecting suitable sample dimensions. The framework also suggests extensions to charge-density-wave composites and vestigial phases. The close competition between nematic and charge-$4e$ superconductivity~\cite{FernandesFu2021}, together with the predicted local condensation of charge-$4e$ order at nematic domain walls~\cite{Hecker2024}, makes controlled bending a promising setting in which to pursue such states. More broadly, the same construction can be explored wherever a spatially varying conjugate field selects distinct configurations of a multi-component order parameter and their interpolation allows additional symmetry breaking.

Gradient quantum matter thus offers a common framework for designing electronic interfaces through the boundary conditions imposed on collective order. Controlled bending connects this framework to accessible experimental geometries, while the resulting interfacial states provide both new forms of order and a probe of competing degrees of freedom within quantum materials.

\begin{acknowledgments}
We acknowledge useful discussions with R. M. Fernandes, R. Nagashima, N. Cheng and Y. Hu.
This work was supported by 
the German Research Foundation TRR 288-422213477 ELASTO-Q-MAT,
 B01 (J.S.) B10 (P.J.W.M.), and grant  SFI-MPS-NFS-00006741-05 from the Simons Foundation (J.S.). This project has received funding from the European Research Council (ERC) under the European Union’s Horizon 2020 research and innovation program XBEND (Grant agreement No. 101080740).
\end{acknowledgments}

\appendix

\makeatletter
\@addtoreset{figure}{section}
\@addtoreset{table}{section}
\makeatother

\renewcommand{\thefigure}{\thesection\arabic{figure}}
\renewcommand{\thetable}{\thesection\arabic{table}}

\twocolumngrid

\section{Bending geometry and finite element simulation}
\label{app:fem}
Fig.~\ref{fig:bendingschematic} schematically illustrates  the geometry of a cantilever of length $D$, width $w$ and thickness $L$ described in Sec.~\ref{sec:geometry}. Bending the cantilever upwards by an angle $\theta$ bends the strain-neutral plane at $z=0$ into an arc segment $s$ with radius of curvature $R$. The top surface of the cantilever then experiences compressive strain, while the bottom surface is subjected to tension, corresponding to a sign-changing strain gradient along the local normal to the strain-neutral plane.\\
A specific example of an experimental geometry implementing controlled bending of single crystals microstructured using focused-ion-beam (FIB) machining is provided by Ref.~\cite{Diaz2022}. The crystals are shaped into cantilevers suspended from the edge of a substrate and subject to bending deformation via a vertical pushing mechanism. Instead of a strip of material of homogeneous thickness, a thinner section acting as a flexure is introduced to the cantilever  (Fig.~\ref{fig:bendingCOMSOL}(a)). When the pushing mechanism is engaged to displace the far end of the sample, the flexure acts as a strain focusing element that provides a homogeneous distribution of the in-plane strain (Fig.~\ref{fig:bendingCOMSOL}(b)). We have performed finite element simulations on this geometry using COMSOL 6.0 to verify the strain distribution within the bent cantilever, demonstrating the presence of a sign-changing strain field with an extended interfacial plane between positive and negative strains.

\begin{figure}[tbp]
    \centering
    \includegraphics[width=\linewidth]{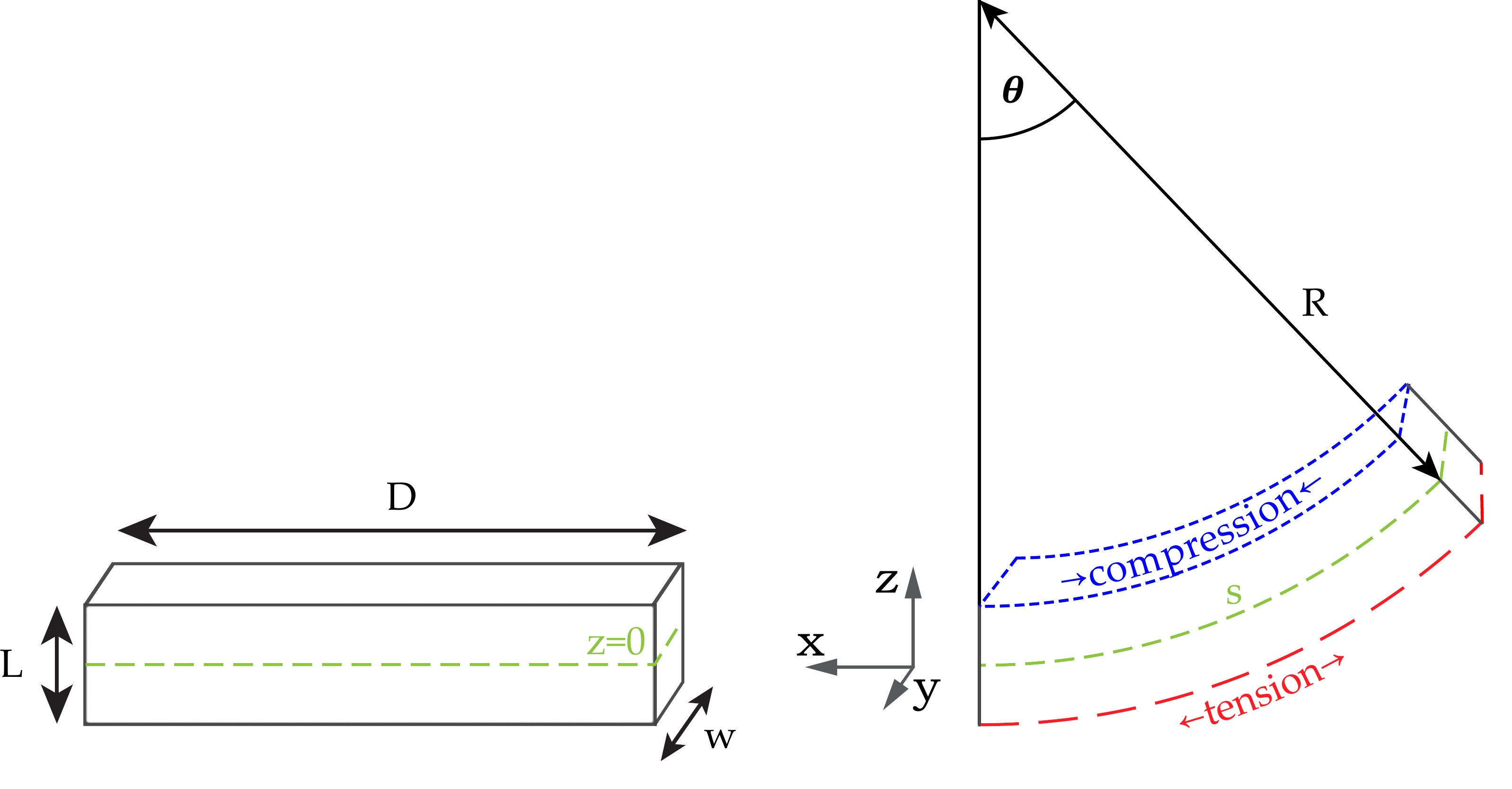}
    \caption{Schematic of the geometry of homogeneous bending. Left: undeformed cantilever of length $D$ along $x$, width $w$ along $y$ and thickness $L$ along $z$. The green dashed line marks the center plane at $z = 0$. Right: Cantilever homogeneously bent around $y$ and subjected to a linear strain gradient from compression at the top (blue) through the strain neutral plane of arc length $s$ (green) to tension at the bottom (red) with radius of curvature $R$ at bending angle $\theta$.}
    \label{fig:bendingschematic}
\end{figure}

\begin{figure}[tbp]
    \centering
    \includegraphics[width=\linewidth]{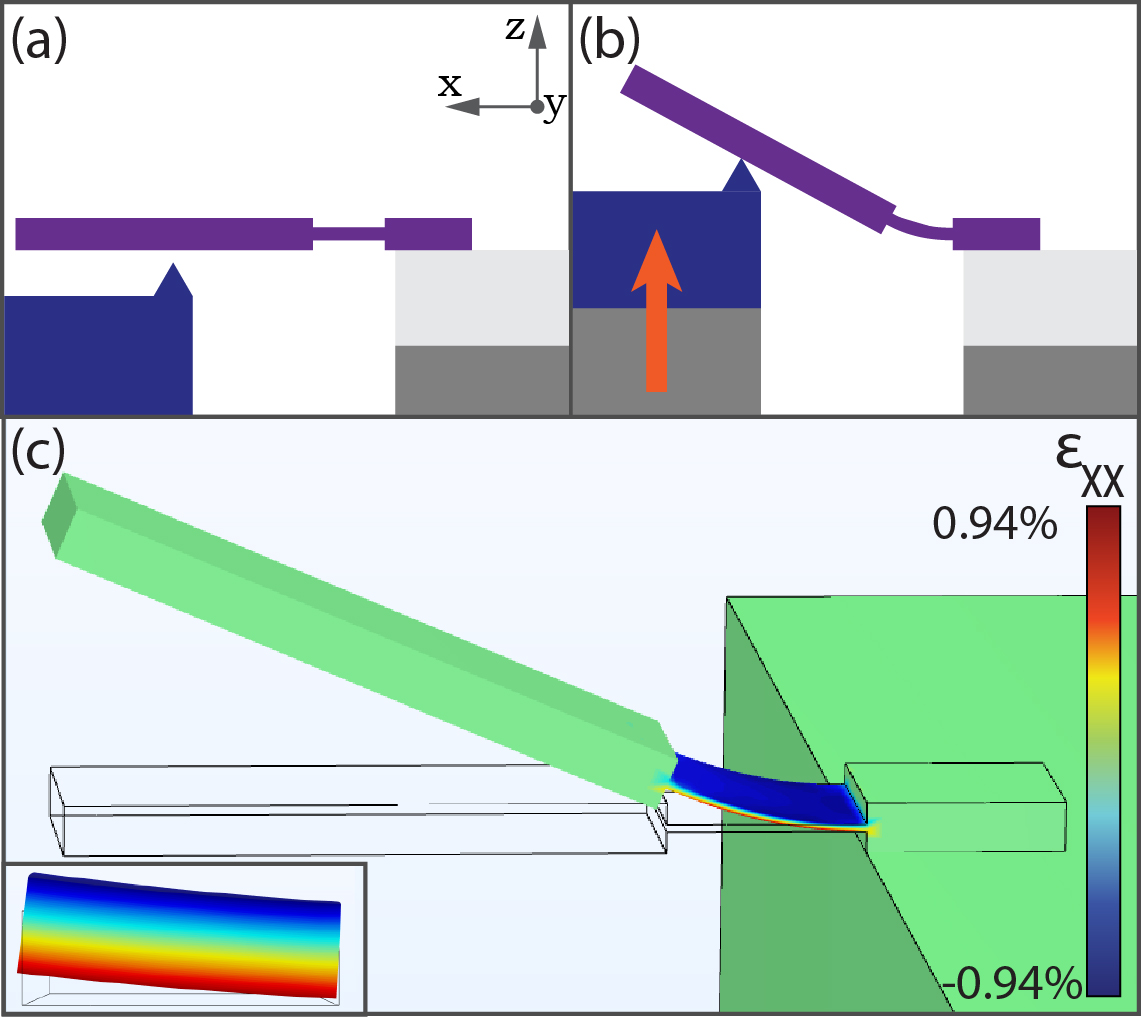}
    \caption{(a,b) Geometry for bending of microstructured single crystals and (c) finite element simulation of the distribution of strain along the length of the cantilever for an isotropic material. Inset shows the strain gradient along a cross-section of the cantilever flexure.}
    \label{fig:bendingCOMSOL}
\end{figure}

\subsection{Finite Element Simulation of Isotropic Material}

For an illustrative isotropic simulation, the material is chosen to be a linear elastic material with a Young's modulus of \SI{30}{\giga\pascal}, corresponding to the stiffness of a prototypical iron-arsenide superconductor \cite{Yoshizawa2012a,Yoshizawa2012b}. The mechanical boundary conditions consist of fixing the sample to the top surface of the substrate, which itself is fixed in space, as is true for the experimental geometry. The vertical displacement of the cantilever is simulated by prescribing a fixed vertical displacement (in the $y$-direction) to the far bottom edge of the cantilever, while the edge is left free to move in the other two dimensions ($x$ and $z$). It is important to leave the paddle of the cantilever free to rotate by not prescribing a displacement in $x$, as this would induce additional uniaxial strain along the length of the cantilever. This mechanical setup corresponds to a sliding contact between the cantilever and the pushing mechanism. In this simulation, a vertical displacement of \SI{60}{\micro\meter} was prescribed, leading to a maximum surface strain of $\pm$\SI{0.94}{\percent}, as illustrated in Fig.~\ref{fig:bendingCOMSOL}(c). The results  show the homogeneity of the in-plane strain, while the inset presents the strain-gradient along the cross-section of the cantilever flexure. These simulations confirm the existence of a strain-neutral plane within the flexure, extending across both its width and its length.

\subsection{Finite Element Simulation of Trigonal Material}
A second simulation was performed for a prototypical trigonal material, Bi$_2$Se$_3$. The isotropic material from the previous simulation was replaced with an anisotropic material, by applying the full elastic tensor as calculated in Ref.\cite{Gao2016}. The material coordinate system is chosen such that the crystallographic a-axis (c-axis) aligns with the $x$-axis ($z$-axis) of the geometry, respectively. The simulation results shown in Fig.~\ref{fig:BiSeSims} confirm that under D$_{3d}$ symmetry, $\varepsilon_{xx}-\varepsilon_{yy}$ transforms as $\varepsilon_{yz}$ with a factor $~0.3-0.4$, and $-\varepsilon_{xy}$ transforms as $\varepsilon_{xz}$ and is of the same magnitude, for both upward and downward bending. For a system with perfect alignment of the a-axis along the geometric $x$-axis, however, a degeneracy arises in the $-\varepsilon_{xy}$ and $\varepsilon_{xz}$ elements (Fig.~\ref{fig:BiSeSims} (d)-(j)). While reversing the directionality of the bending deformation reverses the sign of the strain in the corners, the in-plane degeneracy is not lifted. Breaking of the degeneracy in the planes can be achieved through an in-plane rotation of the crystal a-axis away from the geometric $x$-axis for a fixed direction of bending (Fig.~\ref{fig:BiSeRotSims}(d)-(j)).

\begin{figure}[tbp]
    \centering
    \includegraphics[width=\linewidth]{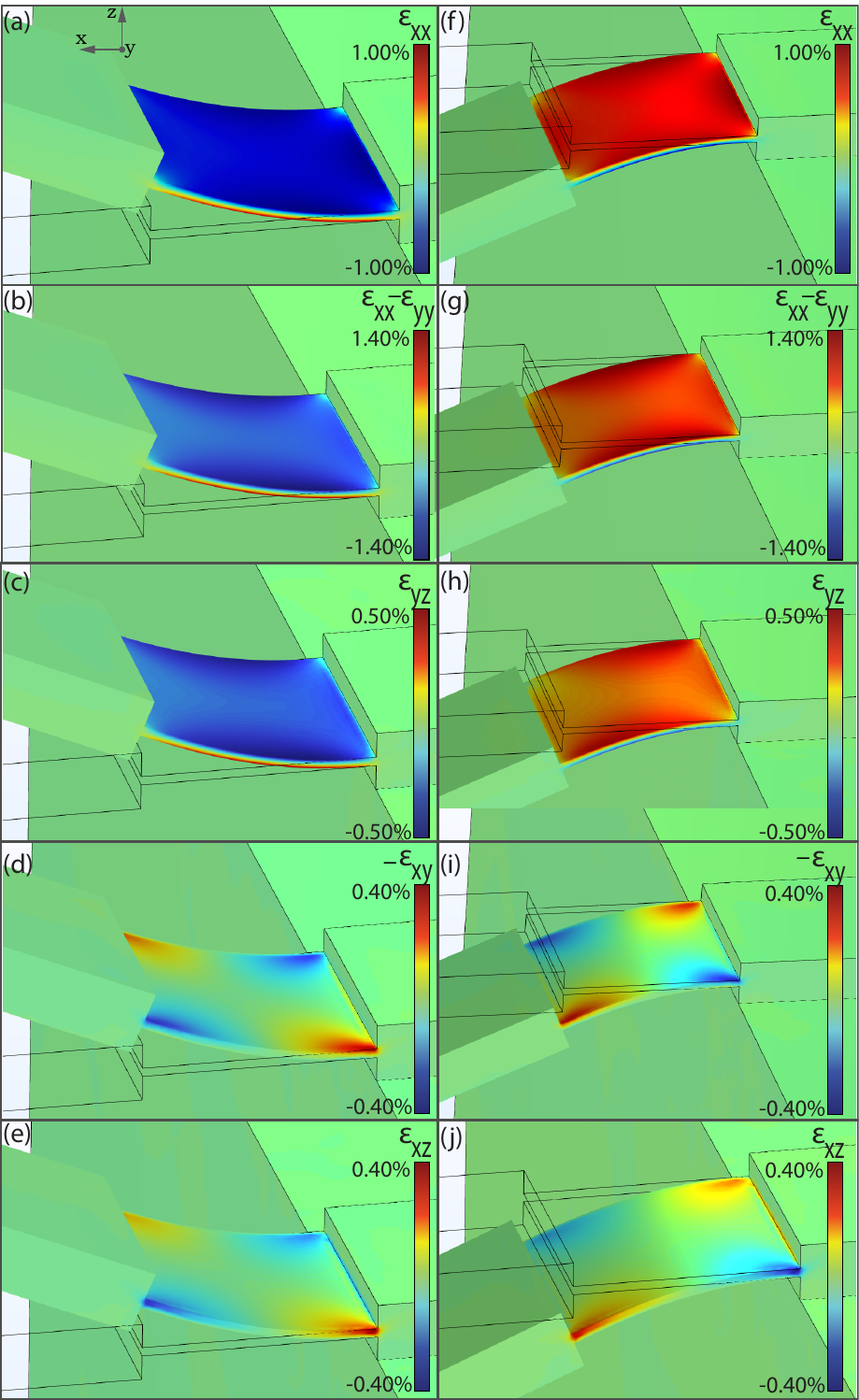}
    \caption{Finite element simulation of the spatial distribution of strain tensor elements resulting from bending a system of trigonal symmetry whose a-axis is along the $x$-direction and whose c-axis is along the $z$-direction. (a)-(e) Bending upwards. (f)-(j) Bending downwards.}
    \label{fig:BiSeSims}
\end{figure}

\begin{figure}[tbp]
    \centering
    \includegraphics[width=\linewidth]{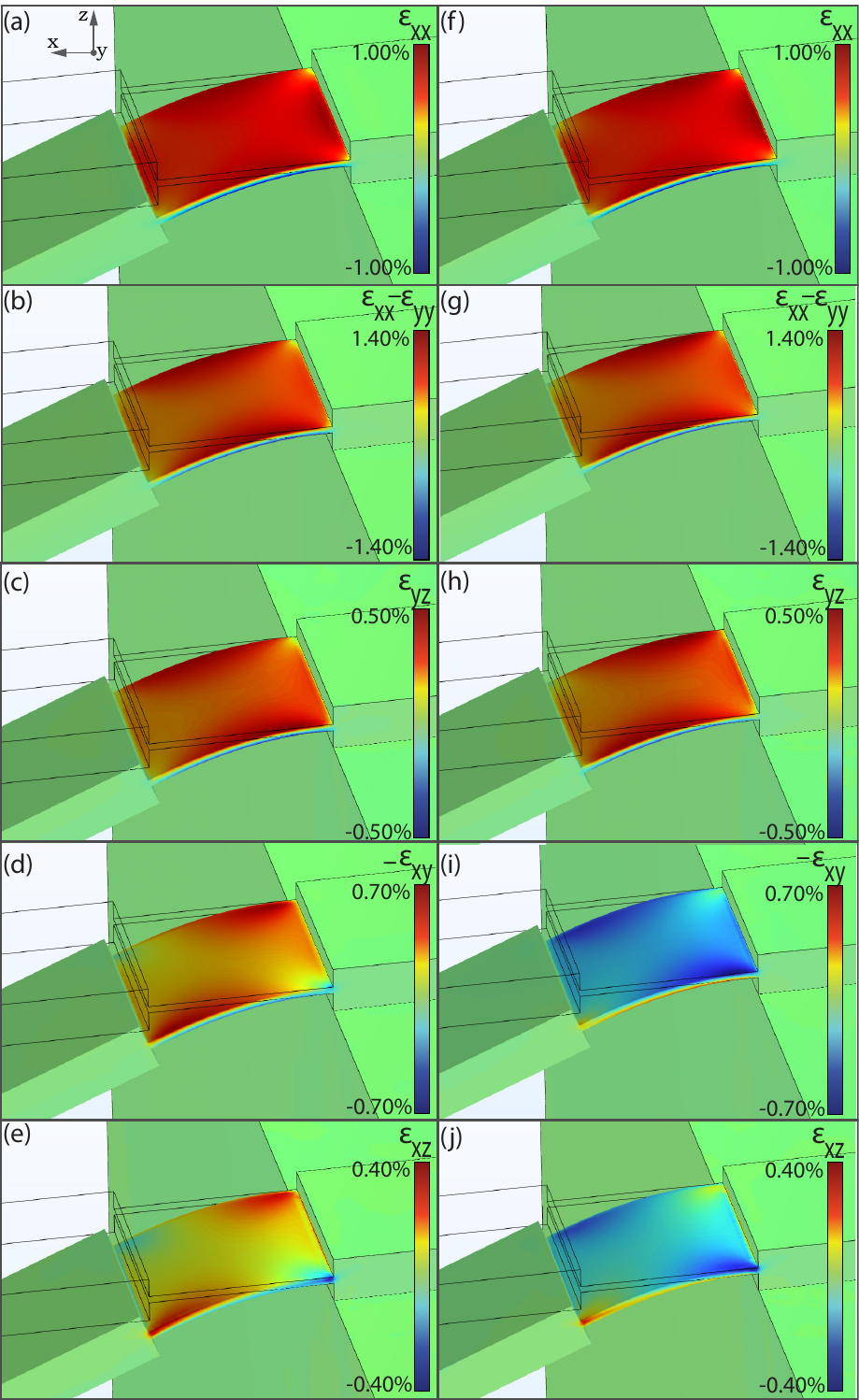}
    \caption{Finite element simulation of the spatial distribution of strain tensor elements resulting from downward bending of a system of trigonal symmetry whose c-axis is along the $z$-direction, but whose a-axis is misaligned by an angle $\theta$ from the geometric $x$-axis, breaking the degeneracy in $-\varepsilon_{xy}$ and $\varepsilon_{xz}$. (a)-(e) $\theta=-\SI{10}{\degree}$. (f)-(j) $\theta=-\SI{10}{\degree}$.}
    \label{fig:BiSeRotSims}
\end{figure}

\section{Ginzburg--Landau theory for stripe magnetism}
\label{app:stripe_gl}

We give details of the GL theory in Sec.~\ref{sec:stripe_magnetism}. Write $\mathbf{S}_X = S_X \hat{n}_X$ and $\mathbf{S}_Y = S_Y \hat{n}_Y$. For collinear case, $\phi = C = \pm S_X S_Y$ and for orthogonal axes in the $xy$ plane, $\phi = V = \pm S_X S_Y$. These two branches therefore have the same amplitude functional 
\begin{equation}
\begin{aligned}
f&=\frac{K_s}{2}\left[(\partial_z S_X)^2+(\partial_z S_Y)^2\right]+\frac r2(S_X^2+S_Y^2)\\
 &+\frac b4(S_X^2+S_Y^2)^2+g_\mathrm{eff}S_X^2S_Y^2-h(z)(S_X^2-S_Y^2).
\end{aligned}
\label{eq:app_mag_0}
\end{equation}
Here $b=u-g>0$ and $w$ refers to the GL coefficient in the selected channel. The bilinears are constrained by $R^2 = N^2 + 4 \phi^2$ on either branch and are not varied independently. We take $g_\mathrm{eff}>0$, ensuring quartic stability and a single-$Q$ preference. These calculations determine the amplitudes within each spin branch.   

\subsection{Choice of units and dimensionless functional }

We write $r = a t$ with $a= a_m T_m>0$ and $t = (T-T_m)/T_m$, and define 
\begin{equation}
\begin{aligned}
\bar S=\sqrt{\frac{2a}{b}},\, \,\bar f=\frac{a^2}{b},\,\, \zeta=\frac zL,\,\, m_{X,Y}=\frac{S_{X,Y}}{\bar S}.
\end{aligned}
\label{eq:app_mag_1}
\end{equation}   
 With $\tilde{f} = f/\bar{f}$ and $\mathcal F=\int d\zeta\,\widetilde f$, we introduce 
\begin{equation}
\begin{aligned}
x_m^2=\frac{K_s}{aL^2},\quad &\kappa=\frac{g_\mathrm{eff}}{b}>0,\quad H(\zeta)=H_\infty\tanh\zeta,
\end{aligned}
\label{eq:app_mag_2}
\end{equation}    
where $g_\mathrm{eff}= g+w$ for the collinear branch and $g_\mathrm{eff} = g$ for the orthogonal branch. Here  $H_\infty={2h_\infty}/a$.

The dimensionless density is 
\begin{equation}
\begin{aligned}
\widetilde f&=x_m^2\left[(m_X')^2+(m_Y')^2\right]+t(m_X^2+m_Y^2)
\\ &+(m_X^2+m_Y^2)^2+4\kappa m_X^2m_Y^2-H(\zeta)(m_X^2-m_Y^2).
\end{aligned}
\label{eq:app_mag_3}
\end{equation}
where a prime denotes $\partial_\zeta$. The dimensionless bilinears follow from 
\begin{equation}
\begin{aligned}
\frac{R}{\bar S^2}=m_X^2+m_Y^2,\quad
&\frac{N}{\bar S^2}=m_X^2-m_Y^2,\quad\frac{\phi}{\bar S^2}=\pm m_Xm_Y.
\end{aligned}
\label{eq:app_mag_4}
\end{equation} 

\subsection{Ginzburg--Landau equations and bulk boundary conditions}

Varying $\mathcal{F}$ with respect to the two amplitudes gives 
\begin{equation}
\begin{aligned}
x_m^2m_X''=(t-H)m_X+2m_X^3+2(1+2\kappa)m_Xm_Y^2.
\end{aligned}
\label{eq:app_mag_mX}
\end{equation}
\begin{equation}
\begin{aligned}
x_m^2m_Y''= (t+H)m_Y+2m_Y^3
 +2(1+2\kappa)m_Ym_X^2.
\end{aligned}
\label{eq:app_mag_mY}
\end{equation} 

The homogeneous minima are single-$Q$: at fixed $R$, the potential is concave in $N$ because $g+w>0$, and its minimum occurs at $N = \pm R$. We focus on the magnetically ordered regime $t<0$, with $H_\infty>0$. The asymptotic boundary conditions are 
\begin{equation}
\begin{aligned}
(m_X,m_Y)\big|_{\zeta\to-\infty}&=(0,m_0),\\
(m_X,m_Y)\big|_{\zeta\to+\infty}&=(m_0,0).
\end{aligned}
\label{eq:app_mag_6}
\end{equation}     
where $m_0^2 = (H_\infty-t)/2$.  

The parameters used in the main text are  
\begin{equation}
\begin{aligned}
H(\zeta)&=H_\infty\tanh\zeta,\\
H_\infty&=\frac{2h_\infty}a=0.01,\qquad \kappa=0.2.
\end{aligned}
\label{eq:app_mag_7}
\end{equation}
The mixing-angle and bilinear profiles use $x_m=0.005$.
The mixing-angle profile is shown at $T/T_m=0.5$, while the bilinear
profiles span $0.5\leq T/T_m\leq0.9$.
For the width scan, $t$ is sampled at 20 linearly spaced values from
$-0.5$ to $-0.1$, and $x_m$ at 20 linearly spaced values from
$10^{-4}$ to $10^{-2}$.

The plotted angle is $\gamma=\tan^{-1}(m_Y/m_X)$.
The bilinear is normalized as $|\phi|/\phi_0=|m_Xm_Y|$, and its width is defined as 
\begin{equation}
\ell_\phi=
 \left[
 \frac{\displaystyle\int dz\,z^2|\phi(z)|}
      {\displaystyle\int dz\,|\phi(z)|}
 \right]^{1/2}.
\label{eq:app_mag_8}
\end{equation}
Reflection symmetry places the centroid at $z=0$; the moments are
evaluated over the full numerical domain.
The black dashed curves use the weak-strain, constant-amplitude
approximation at $T/T_m=0.5$.

\subsection{Analytical profile of the double-$Q$ order $\phi$}

To describe the interpolation from one domain to another, we introduce a total amplitude $A$ and a mixing angle $\gamma$,
\begin{equation}
m_X = A \cos \gamma,\qquad m_Y = A \sin \gamma.
\label{eq:app_mag_9}
\end{equation}  

The angle describes the relative weights of the two ordering wavevectors and the dimensionless bilinears are 
\begin{equation}
\begin{aligned}
\frac{R}{\bar S^2}=A^2,\quad &\frac{N}{\bar S^2}=A^2\cos(2\gamma),\quad\frac{\phi}{\bar S^2}=\pm\frac{A^2}{2}\sin(2\gamma).
\end{aligned}
\label{eq:app_mag_10}
\end{equation}

In this notation, the GL functional is 
\begin{equation}
\begin{aligned}
\widetilde f={}&x_m^2\left[(A')^2+A^2(\gamma')^2\right]+tA^2+A^4
\\ &+\kappa A^4\sin^2(2\gamma)-HA^2\cos(2\gamma).
\end{aligned}
\label{eq:app_mag_11}
\end{equation}

The GL equations are 
\begin{equation}
\begin{aligned}
x_m^2\left[A''-A(\gamma')^2\right]&=(t-H\cos2\gamma)A
\\ &+2A^3\left[1+\kappa\sin^2(2\gamma)\right].
\end{aligned}
\label{eq:app_mag_A}
\end{equation}
\begin{equation}
x_m^2(A^2\gamma')'=\kappa A^4\sin(4\gamma)+HA^2\sin(2\gamma).
\label{eq:app_mag_gamma}
\end{equation}

To obtain an analytical wall profile, we assume weak strain within the wall core and weak competing-order anisotropy, $\kappa=g_{\mathrm{eff}}/(u-g)\ll1$. The latter condition makes amplitude relaxation parametrically small, allowing the total magnetic amplitude to be treated as constant to leading order: 
\begin{equation}
A\simeq m_0,\qquad m_0=\sqrt{\frac{H_\infty-t}{2}}\simeq\sqrt{-\frac t2}.
\label{eq:app_mag_13}
\end{equation}

Neglecting $H$ within the wall core, the angular equation reduces to 
\begin{equation}
x_m^2 \gamma'' = \kappa m_0^2 \sin 4 \gamma
\label{eq:app_mag_14}
\end{equation} 

Multiplying by $\gamma'$ and integrating gives 
\begin{equation}
x_m^2 (\gamma')^2 = \kappa m_0^2 \sin^2 2 \gamma
\label{eq:app_mag_15}
\end{equation} 

The boundary conditions $\gamma$ is $\gamma(-\infty) = \pi/2,\gamma(\infty) = 0$  select the decreasing
solution
\begin{equation}
\gamma' = -\frac{\sqrt{\kappa} m_0}{x_m} \sin 2 \gamma.
\label{eq:app_mag_16}
\end{equation}

Therefore the solution is 
\begin{equation}
\begin{aligned}
\cos(2\gamma)&=\tanh\!\left(\frac{\zeta}{\widetilde\ell_w}\right),
\end{aligned}
\label{eq:app_mag_17}
\end{equation}
where $\widetilde\ell_w = {x_m}/{\sqrt{2\kappa(H_\infty-t)}}$.

In dimensional units, the weak-strain bilinear profile is 
\begin{equation}
\phi(z)=\pm\frac{a}{2b}\frac{T_m-T}{T_m}
\operatorname{sech}\!\left(\frac z{\ell_w}\right).
\label{eq:app_mag_18}
\end{equation}
where $\ell_w$ is the dimensionful sech-profile decay length. In the weak-strain limit, 
\begin{equation}
\ell_w=L\widetilde\ell_w\simeq\frac{\xi_m}{\sqrt{-2\kappa t}},
\qquad \xi_m=\sqrt{\frac{K_s}{a}}.
\label{eq:app_mag_19}
\end{equation}

The second-moment width used in the main text is distinct from this
profile length. For the sech profile, it is
\begin{equation}
\ell_\phi=\frac\pi2\ell_w
\simeq\frac{\pi\xi_m}{\sqrt{-8\kappa t}}.
\label{eq:app_mag_rms_width}
\end{equation}
Thus the same profile can be written with argument $\pi z/(2\ell_\phi)$. 

\section{Ginzburg--Landau theory for tetragonal nematic superconductivity}
\label{app:tetragonal_gl}

\label{sec:tet_dimensionless}

We formulate the tetragonal domain-wall problem in dimensionless variables
using the same normalization of the isotropic condensation energy as for
the trigonal model. The superconducting amplitude remains a variational
field throughout. We consider a texture that varies only along $z$ and
write $F=\int dz\,f$ for the free energy per transverse area. In the absence
of an imposed transport current or an externally fixed phase difference,
the common phase obeys
\begin{equation}
\partial_z\varphi+\frac12\sin\eta\,\partial_z\beta=0.
\label{eq:tet_zero_current}
\end{equation}
Eliminating this phase gives the reduced dimensional energy density
\begin{equation}
\begin{aligned}
f&=\frac d2(\partial_z\rho)^2
+\frac{d\rho^2}{8}\left[(\partial_z\eta)^2
+\cos^2\eta(\partial_z\beta)^2\right]
\\ &+\frac r2\rho^2+\frac u4\rho^4
+\frac{u''\rho^4}{16}\cos^2\eta\sin^2\beta
\\ &+\frac{(u''-4u')\rho^4}{16}\sin^2\eta
-H(z)\rho^2\cos\eta\cos\beta.
\end{aligned}
\label{eq:tet_dimensional_density}
\end{equation}
where $H(z)=\lambda\varepsilon(z)$ includes the coupling to the imposed
$B_{1g}$ strain. The symbols $u'$ and $u''$ denote quartic coefficients.
We take $d,u>0$ and $u''>\max(4u',0)$, so that the unstrained homogeneous
superconducting minima are the two $B_{1g}$ nematic states.

\subsection{Choice of units and dimensionless functional}
\label{subsec:tet_units}

We write
\begin{equation}
r=at,\qquad t=\frac{T-T_c}{T_c},\qquad a>0,
\label{eq:tet_temperature}
\end{equation}
where $T_c$ is the zero-strain mean-field transition temperature. For a
regularized sign-changing strain profile, we choose
\begin{equation}
H(z)=H_\infty \tanh(z/L),\qquad H_\infty >0,
\label{eq:tet_strain}
\end{equation}
with $L$ the strain-variation length. This profile is linear near the
neutral plane and saturates in the asymptotic domains. The same scaling
applies to other prescribed strain profiles.

The amplitude and energy-density units are
\begin{equation}
\rho_0=\sqrt{\frac{2a}{u}},\qquad f_0=\frac{a^2}{u},
\label{eq:tet_units}
\end{equation}
and we introduce
\begin{equation}
\begin{aligned}
\zeta=\frac zL,\qquad &\widetilde\rho=\frac\rho{\rho_0},
\qquad \widetilde f=\frac f{f_0}.
\end{aligned}
\label{eq:tet_scaled_fields}
\end{equation}
Here $\rho_0$ is a fixed normalization, rather than the equilibrium
amplitude at a particular temperature. The dimensionless coefficients are
\begin{equation}
\begin{aligned}
x_s^2=\frac{d}{aL^2},\quad &\kappa_\beta=\frac{u''}{4u},\quad\kappa_\chi=\frac{u''-4u'}{4u}.\\
\end{aligned}
\label{eq:tet_dimensionless_parameters}
\end{equation}

The dimensionless strain is 
\begin{equation}
\begin{aligned}
h(\zeta)=\frac{2H(L\zeta)}a&=h_\infty\tanh\zeta,\\
\end{aligned}
\label{eq:tet_dimensionless_parameters_st}
\end{equation}
where $ h_\infty={2H_\infty}/a$.

Dropping the tilde on the amplitude and using a prime for a derivative
with respect to $\zeta$, the dimensionless energy density becomes
\begin{equation}
\begin{aligned}
\widetilde f={}&x_s^2\left[(\rho')^2
+\frac{\rho^2}{4}\left((\eta')^2+\cos^2\eta(\beta')^2\right)\right]
\\ &+t\rho^2+\rho^4
+\rho^4\left[\kappa_\beta\cos^2\eta\sin^2\beta
+\kappa_\chi\sin^2\eta\right]
\\ &-h(\zeta)\rho^2\cos\eta\cos\beta.
\end{aligned}
\label{eq:tet_dimensionless_density}
\end{equation}
 
The dimensionless bilinears are
\begin{equation}
\begin{aligned}
(N_1,N_2)&=\rho^2\cos\eta(\cos\beta,\sin\beta),\quad 
\chi=\rho^2\sin\eta.
\end{aligned}
\label{eq:tet_dimensionless_bilinears}
\end{equation}
so the quartic anisotropy is simply
$\kappa_\beta N_2^2+\kappa_\chi\chi^2$. The dimensional bilinears are
obtained by multiplying these expressions by $\rho_0^2$.
The $B_{1g}$ nematic regime corresponds to
$\kappa_\beta,\kappa_\chi>0$ and within this regime, $u'>0$ is equivalent to
$0<\kappa_\chi<\kappa_\beta$. 

\subsection{Ginzburg--Landau equations and bulk boundary conditions}
\label{subsec:tet_scaled_equations}

Variation of $\mathcal F$ with respect to $\rho$, $\beta$, and $\eta$ gives
\begin{equation}
\begin{aligned}
x_s^2\rho''&=\frac{x_s^2\rho}{4}
\left[(\eta')^2+\cos^2\eta(\beta')^2\right]+(t-h\cos\eta\cos\beta)\rho
\\ &+2\rho^3\left[1+\kappa_\beta\cos^2\eta\sin^2\beta
+\kappa_\chi\sin^2\eta\right].
\end{aligned}
\label{eq:tet_scaled_rho}
\end{equation}
\begin{equation}
\begin{aligned}
x_s^2(\rho^2\cos^2\eta\,\beta')'&=
2\kappa_\beta\rho^4\cos^2\eta\sin(2\beta)
\\ &+2h\rho^2\cos\eta\sin\beta.
\end{aligned}
\label{eq:tet_scaled_beta}
\end{equation}
\begin{equation}
\begin{aligned}
x_s^2(\rho^2\eta')'&=-\frac{x_s^2\rho^2}{2}
\sin(2\eta)(\beta')^2+2h\rho^2\sin\eta\cos\beta
\\ &+2\rho^4\sin(2\eta)
(\kappa_\chi-\kappa_\beta\sin^2\beta).
\end{aligned}
\label{eq:tet_scaled_eta}
\end{equation}

The bulk boundary values follow by minimizing the homogeneous potential
at the prescribed endpoint $\zeta = \pm \infty$  with strain $\pm h_\infty$ . Writing $n=\rho^2$, for the symmetric profile in Eq.~\eqref{eq:tet_strain}, the asymptotic
superconducting boundary conditions are therefore
\begin{equation}
\begin{aligned}
&\rho(\pm\infty)=\sqrt{\frac{h_\infty-t}{2}},\quad
\eta(\pm\infty)=0,\\
&\beta(-\infty)=\pi,\quad \beta(+\infty)=0.
\end{aligned}
\label{eq:tet_scaled_infinite_bc}
\end{equation}
provided $t<0<h_\infty$. In particular, the zero-strain bulk amplitude at $t<0$
is $\rho^2(\pm \infty)=-t/2$.  

We solve the stationary boundary-value equations by adaptive collocation, with the same bulk boundary conditions
for both TRS and TRSB branches. The TRS solution is obtained by imposing $\eta=0$
and solving the coupled equations for $\rho$ and $\beta$. To search for
a TRSB solution, we initialize these two fields with their TRS profiles
and introduce the localized seed
\begin{equation}
\begin{aligned}
\eta_{\mathrm{init}}(\zeta)&=\eta_0 \left(1-\frac{\zeta^2}{\Lambda^2}\right)e^{-\zeta^2/w^2}
\end{aligned}
\label{eq:tet_numerical_seed}
\end{equation}
where   $0<|\eta_0|<\frac\pi2, w>0$.

\begin{table}[tb]
\centering
\label{tab:app_tetragonal_parameters}
\small
\setlength{\tabcolsep}{4pt}
\begin{tabular}{@{}lclc@{}}
\toprule
Parameter & Value & Parameter & Value \\
\midrule
$t$ & $[-0.5,0]$ & $x_s$ & $0.01$ \\
$\kappa_\beta$ & $0.1$ & $\kappa_\chi$ & $0.05$ \\
$h_\infty$ & $0.01$ & & \\
\bottomrule
\end{tabular}
\caption{Parameters for the tetragonal textures in Fig.~\ref{fig:tetra_SC}.}
\end{table}

Then solve all three coupled equations without constraining the interior
$\eta$ profile. A converged solution with nonzero
$\chi=\rho^2\sin\eta$ defines a TRSB branch, whose time-reversed
partner has the same energy. We evaluate the full functional on each converged profile. The dimensionless excess energy is obtained by subtracting the same bulk reference for both branches: 
\begin{equation}
\begin{aligned}
\mathcal F_s^{\mathrm{ex}}&=\int_{-\Lambda}^{\Lambda}d\zeta\,
\left[\widetilde f_s(\zeta)-\widetilde f_{\mathrm{bulk}}\right], 
\end{aligned}
\end{equation}
where $s\in\{\mathrm{TRS},\mathrm{TRSB}\}$ and $\widetilde f_{\mathrm{bulk}}$ is the constant endpoint bulk energy density.
The numerical parameters are summarized in Table~\ref{tab:app_tetragonal_parameters}. 
 
\subsection{Analytical profile of a tetragonal TRSB domain wall}
\label{sec:tet_analytic_wall}

We derive the analytical wall profile in the regime
$d,u>0$ and $0<u'<u''/4$, for which the homogeneous superconducting
minima are nematic while a complex interfacial texture can lower the
energy. We use dimensional fields and coefficients in the derivation
below and write $F=\int dz\,f$ for the free energy per unit transverse
area. The strain coupling is $-H(z)(|\psi_1|^2-|\psi_2|^2)$, with
$H(z)=H_\infty\tanh(z/L)$ and $H_\infty>0$. Thus the left and right bulk domains
contain only $\psi_2$ and $\psi_1$, respectively. Their superconducting
phases are free, and no transport current is imposed.

For the analytical profile, approximate the amplitude by a constant
$\bar\rho>0$. It is more convenient to use the following  parametrization
\begin{equation}
\begin{aligned}
\Psi(z)&=\bar\rho e^{i\alpha_0}
\begin{pmatrix}\cos\theta(z)\\ e^{i\delta}\sin\theta(z)\end{pmatrix},\,0\leq\theta\leq\frac\pi2.
\end{aligned}
\label{eq:tet_wall_mixing_angle}
\end{equation}
Here $\alpha_0$ and $ \delta$ are constant phases, and $\theta$ is a
component-mixing angle.
In the weak-strain bulk approximation, $\bar\rho^2\simeq-r/u$ for $r<0$.

After dropping the constant isotropic condensation energy
$r\bar\rho^2/2+u\bar\rho^4/4$, the remaining energy density is
\begin{equation}
\begin{aligned}
f_r&=\frac{d\bar\rho^2}{2}(\partial_z\theta)^2
+\frac{\bar\rho^4A_\delta}{4}\sin^2(2\theta)-H(z)\bar\rho^2\cos(2\theta),
\end{aligned}
\label{eq:tet_wall_reduced_energy}
\end{equation}
where $A_\delta={u''}/{4}-u'\sin^2\delta$. 

For the two branches,
\begin{equation}
\begin{aligned}
A_{\mathrm{TRS}}=\frac{u''}{4},\qquad
&A_{\mathrm{TRSB}}=\frac{u''}{4}-u'.\\
\end{aligned}
\label{eq:tet_wall_anisotropies}
\end{equation}
Here we have $0<A_{\mathrm{TRSB}}<A_{\mathrm{TRS}}$. 

The boundary conditions are
\begin{equation}
\begin{aligned}
\theta(-\infty)=\frac\pi2,\quad &\theta(+\infty)=0, \quad 
 \partial_z\theta(\pm\infty)=0.
\end{aligned}
\label{eq:tet_wall_theta_bc}
\end{equation}

We now take the limit in which strain selects the asymptotic domains
but gives a small correction to the wall core. Neglecting the strain
term in the local wall equation gives
\begin{equation}
d\bar\rho^2\partial_z^2\theta
 =\frac{\bar\rho^4  A_\delta}{2}\sin(4\theta).
\label{eq:tet_wall_theta_el}
\end{equation}
Multiplication by $\partial_z\theta$ and integration yield
\begin{equation}
\frac{d\bar\rho^2}{2}(\partial_z\theta)^2
 -\frac{\bar\rho^4  A_\delta}{4}\sin^2(2\theta)=0,
\label{eq:tet_wall_first_integral}
\end{equation}
where the integration constant vanishes by the bulk boundary conditions.
Choosing the decreasing solution, we have 
\begin{equation}
\partial_z\theta=-\frac{\sin(2\theta)}{2\ell_\chi},\qquad
\ell_\chi=\sqrt{\frac{d}{2\bar\rho^2A_\delta}}.
\label{eq:tet_wall_width}
\end{equation}
Since $\partial_z\ln(\tan\theta)
=2\,\partial_z\theta/\sin(2\theta)$, the solution is
\begin{equation}
\begin{aligned}
\tan\theta(z)&=e^{-z/\ell_\chi}.\\
\end{aligned}
\label{eq:tet_wall_kink}
\end{equation}
We have placed the wall center at $z=0$, as selected by the
reflection-symmetric strain profile. Without strain, its position is
an arbitrary integration constant.

For $\delta = \pm\pi/2$, the analytical TRSB texture is
\begin{equation}
\begin{aligned}
\Psi_\pm(z)&=\frac{\bar\rho e^{i\alpha_0}}{\sqrt{1+e^{-2z/\ell_\chi}}}
\begin{pmatrix}1\\ \pm i e^{-z/\ell_\chi}\end{pmatrix},
\end{aligned}
\label{eq:tet_wall_spinor}
\end{equation}
where $\ell_\chi=\sqrt{{d}/[{2\bar\rho^2(u''/4-u')]}}$. 

The nematic bilinears are 
\begin{equation}
\begin{aligned}
N_1(z)&=\bar\rho^2\tanh(z/\ell_\chi),\qquad N_2(z)=0. 
\end{aligned}
\label{eq:tet_wall_neamtic}
\end{equation}

The axial bilinear is 
\begin{equation}
\begin{aligned}
\chi(z)&=\pm\bar\rho^2\operatorname{sech}(z/\ell_\chi).
\end{aligned}
\label{eq:tet_wall_chirality}
\end{equation}

The two signs are related by time reversal. Although the relative phase
is constant throughout the ansatz, the axial order parameter vanishes in the bulk because
only one component survives there. At the wall center the components have
equal magnitude, giving $|\chi(0)|=\bar\rho^2$ and $\eta(0)=\pm\pi/2$.
More generally,
$\sin\eta=\pm\operatorname{sech}(z/\ell_\chi)$ and
$\cos\eta\cos\beta=\tanh(z/\ell_\chi)$.
Thus $\beta=\pi$ for $z<0$ and $\beta=0$ for $z>0$, while $\beta$ is
undefined at the pole $\eta=\pm\pi/2$, reached at $z=0$. This coordinate change introduces
no discontinuity in the order parameter or its bilinears.

\section{Ginzburg--Landau theory for trigonal nematic superconductivity}
\label{app:trigonal_gl}
\label{sec:tri_supp}

We use the trigonal free-energy density of the main text and the same
normalization as in the tetragonal section. The texture varies only along
the trigonal axis $z$, and $F=\int dz\,f$ denotes the free energy per unit
transverse area. With no imposed transport current or fixed phase difference
between the bulk domains, the common superconducting phase satisfies
\begin{equation}
\partial_z\varphi+\frac12\sin\eta\,\partial_z\beta=0.
\label{eq:tri_supp_current}
\end{equation}
Eliminating this phase gives
\begin{equation}
\begin{aligned}
f={}&\frac d2(\partial_z\rho)^2
+\frac{d\rho^2}{8}\left[(\partial_z\eta)^2
+\cos^2\eta(\partial_z\beta)^2\right]
\\ &+\frac r2\rho^2+\frac u4\rho^4+\frac g4\rho^4\sin^2\eta
\\ &+\frac v6\rho^6(1-\cos^3\eta\cos3\beta)
-\rho^2\cos\eta\,h_\parallel(z,\beta).
\end{aligned}
\label{eq:tri_supp_dimensional}
\end{equation}
where $d,u,g,v>0$ and
$h_\parallel(z,\beta)=h_\infty\tanh(z/L)\cos(\beta-\alpha)$.

\subsection{Choice of units and dimensionless functional}
\label{subsec:tri_supp_units}

As in the tetragonal problem, we write
$r=at$, where $t=(T-T_c)/T_c$, $a>0$, and $T_c$ is the zero-strain
mean-field transition temperature. We define
\begin{equation}
\begin{aligned}
&\rho_0=\sqrt{\frac{2a}{u}},\qquad f_0=\frac{a^2}{u},\\
&\zeta=\frac zL,\qquad \widetilde\rho=\frac\rho{\rho_0},
\qquad \widetilde f=\frac f{f_0},\\
&\mathcal F=\frac{F}{f_0L}=\int d\zeta\,\widetilde f.
\end{aligned}
\label{eq:tri_supp_units}
\end{equation}
The scale $\rho_0$ is temperature independent and is not the equilibrium
amplitude at a particular temperature. The dimensionless coefficients are
\begin{equation}
\begin{aligned}
x_s^2=\frac{d}{aL^2},\qquad &\widetilde g=\frac gu,
\qquad \widetilde v=\frac{4av}{3u^2}. 
\end{aligned}
\end{equation}

The dimensionless strain field is 
\begin{equation}
\begin{aligned}
\widetilde{\mathbf h}(\zeta)&=\widetilde h_\infty\tanh\zeta
(\cos\alpha,\sin\alpha)
\end{aligned}
\end{equation}
where $ \widetilde h_\infty={2h_\infty}/a$.
 
We also define
$\tilde h_\parallel=\tilde h_1\cos\beta+
\tilde h_2\sin\beta$ and
$\tilde h_\perp=\partial_\beta\tilde h_\parallel
=-\tilde h_1\sin\beta+\tilde h_2\cos\beta$.
Below we drop the tildes on the dimensionless fields and coefficients, retaining $\widetilde f$ for the density; a prime denotes
$\partial_\zeta$. The reduced density is
\begin{equation}
\begin{aligned}
\widetilde f={}&x_s^2\left[(\rho')^2
+\frac{\rho^2}{4}\left((\eta')^2+\cos^2\eta(\beta')^2\right)\right]
\\ &+t\rho^2+\rho^4+g\rho^4\sin^2\eta
\\ &+v\rho^6(1-\cos^3\eta\cos3\beta)
-\rho^2\cos\eta\,h_\parallel(\zeta,\beta).
\end{aligned}
\label{eq:tri_supp_density}
\end{equation}
 
The dimensionless bilinears are
\begin{equation}
\begin{aligned}
\mathbf{N}&=\rho^2\cos\eta(\cos\beta,\sin\beta),\\
\chi&=\rho^2\sin\eta,\qquad |\mathbf{N}|^2+\chi^2=\rho^4.
\end{aligned}
\label{eq:tri_supp_bilinears}
\end{equation}
Multiplication by $\rho_0^2$ restores their dimensional values.

\subsection{Ginzburg--Landau equations and bulk boundary conditions}
\label{subsec:tri_supp_equations}

Variation of Eq.~\eqref{eq:tri_supp_density} gives
\begin{equation}
\begin{aligned}
x_s^2\rho''&=\frac{x_s^2\rho}{4}
\left[(\eta')^2+\cos^2\eta(\beta')^2\right]
 +(t-h_\parallel\cos\eta)\rho
\\ &+2\rho^3(1+g\sin^2\eta)
 +3v\rho^5(1-\cos^3\eta\cos3\beta).
\end{aligned}
\label{eq:tri_supp_rho}
\end{equation}
\begin{equation}
\begin{aligned}
x_s^2(\rho^2\cos^2\eta\,\beta')' =
6v\rho^6\cos^3\eta\sin3\beta
 -2\rho^2\cos\eta\,h_\perp.
\end{aligned}
\label{eq:tri_supp_beta}
\end{equation}
\begin{equation}
\begin{aligned}
x_s^2(\rho^2\eta')'&=-\frac{x_s^2\rho^2}{2}
\sin(2\eta)(\beta')^2
 +2g\rho^4\sin(2\eta)
\\ &+6v\rho^6\cos^2\eta\sin\eta\cos3\beta
 +2\rho^2\sin\eta\,h_\parallel.
\end{aligned}
\label{eq:tri_supp_eta}
\end{equation}
The TRS restriction $\eta=0$ satisfies
Eq.~\eqref{eq:tri_supp_eta}. For a TRSB solution all three
fields must be varied. The angular coordinate $\beta$ becomes redundant
when $|\eta|=\pi/2$, where $\mathbf{N}=0$.

For fixed $n=\rho^2>0$, the allowed nematic bilinears fill the disk
$N_1^2+N_2^2\leq n^2$, with
$\chi^2=n^2-N_1^2-N_2^2$. The homogeneous potential can therefore
be written as
\begin{equation}
\begin{aligned}
\mathcal V_n(\mathbf N)={}&tn+(1+g)n^2+vn^3-g(N_1^2+N_2^2)
\\ &-v(N_1^3-3N_1N_2^2)-\mathbf h\cdot\mathbf N.
\end{aligned}
\label{eq:tri_supp_bulk_disk}
\end{equation}
Its two-dimensional Laplacian is
\begin{equation}
\left(\partial_{N_1}^2+\partial_{N_2}^2\right)\mathcal V_n
 =-4  g<0.
\label{eq:tri_supp_bulk_laplacian}
\end{equation}
An interior local minimum would have a positive-semidefinite Hessian
and hence a nonnegative trace, contradicting
Eq.~\eqref{eq:tri_supp_bulk_laplacian}. The global minimum on the disk
must therefore lie on its boundary, where $|\mathbf{N}|=n$ and
$\chi=0$. Thus every nonzero homogeneous ground state of this minimal
model is TRS, including at finite strain.  

It is consequently sufficient to minimize the homogeneous potential
within $\eta=0$ at $\zeta = \pm \infty$ to get the boundary condition:
\begin{equation}
\begin{aligned}
\mathcal V_\pm(n_\pm,\beta_\pm)={}&
\left[t\mp h_\infty\cos(\alpha-\beta_\pm)\right]n_\pm
\\ &+n_\pm^2+vn_\pm^3(1-\cos3\beta_\pm).
\end{aligned}
\label{eq:tri_supp_bulk_potential}
\end{equation}

For a fixed angle define
$A_\pm =3  v(1-\cos3\beta_\pm )\geq0$ and
$C_\pm =t \mp  h_\infty\cos(\alpha - \beta_\pm)$.
The radial stationarity condition is $An^2+2n+C=0$, and the
nonnegative radial minimum is
\begin{equation}
\begin{aligned}
n_\pm =
 \begin{cases}
 0,&C_\pm \geq0,\\[2pt]
 \displaystyle\frac{-C_\pm }{1+\sqrt{1-A_\pm C_\pm }},&C_\pm <0.
 \end{cases}
\end{aligned}
\label{eq:tri_supp_radial_minimum}
\end{equation}

The bulk orientation stationary equation gives 
\begin{equation}
3v n_{\pm}^{\,2}\sin3\beta_{\pm }
 = \pm h_\infty \sin(\alpha - \beta_\pm).
\label{eq:tri_supp_bulk_angle}
\end{equation}
All stationary solutions $(n_\pm,\beta_\pm)$ must be compared by their full homogeneous free energy to identify the global minimum. 

The parameters of the bulk-angle scan are given in Table~\ref{tab:app_trigonal_bulk_parameters}. 
\begin{table}[htb]
\centering
\label{tab:app_trigonal_bulk_parameters}
\small
\setlength{\tabcolsep}{4pt}
\begin{tabular}{@{}lclc@{}}
\toprule
Parameter & Value & Parameter & Value \\
\midrule
$t$ & $-0.1$ & $v$ & $0.1$ \\
$g$ & $0.01$ & $\alpha$ & $[0,2\pi]$ \\
$h_\infty$ & $10^{-4}$ & & \\
\bottomrule
\end{tabular}
\caption{Parameters for the bulk orientations in Fig.~\ref{fig:tri_bulk}.}
\end{table}

At $\zeta\to\pm\infty$, the boundary conditions are
\begin{equation}
\begin{aligned}
\rho(\pm\infty)=\sqrt{n_\pm},\quad \beta(\pm\infty)=\beta_\pm, \quad \eta(\pm\infty)=0.
\end{aligned}
\label{eq:tri_supp_boundary}
\end{equation}
with vanishing derivatives in the asymptotic domains.  

As in the tetragonal case, we solve both for TRS and TRSB states and compare the excess energy. In the trigonal system, the endpoint energies need not be equal for a general strain direction $\alpha$. We define the dimensionless excess energy by subtracting
the left and right bulk reference on their respective half-intervals:
\begin{equation}
\begin{aligned}
\mathcal F_s^{\mathrm{ex}}={}&\int_{-\Lambda}^{0}d\zeta\,
(\widetilde f_s-\widetilde f_-) +\int_{0}^{\Lambda}d\zeta\,(\widetilde f_s-\widetilde f_+),
\end{aligned}
\label{eq:tri_supp_excess}
\end{equation}
where $s\in\{\mathrm{TRS},\mathrm{TRSB}\}$.

The energy difference between two configurations is 
\begin{equation}
\Delta\mathcal F=\mathcal F_{\mathrm{TRSB}}^{\mathrm{ex}}-\mathcal F_{\mathrm{TRS}}^{\mathrm{ex}}.
\label{eq:tri_supp_deltaF}
\end{equation}
A negative $\Delta\mathcal F$ favors the TRSB branch.   The relative difference
$\Delta\mathcal F/\mathcal F_{\mathrm{TRS}}^{\mathrm{ex}}$ is the
dimensionless quantity shown in the main text.

The parameters of the interfacial textures are given in Table~\ref{tab:app_trigonal_wall_parameters}. 
\begin{table}[tb]
\centering
\label{tab:app_trigonal_wall_parameters}
\small
\setlength{\tabcolsep}{4pt}
\begin{tabular}{@{}lclc@{}}
\toprule
Parameter & Value & Parameter & Value \\
\midrule
$t$ & $[-0.8,0]$ & $v$ & $0.2$ \\
$g$ & $0.01$ & $h_\infty$ & $10^{-3}$ \\
$\alpha$ & $-\pi/10$ & $x_s$ & $0.01$ \\
\bottomrule
\end{tabular}
\caption{Parameters for the trigonal textures in Fig.~\ref{fig:tri_SC}.}
\end{table}

\subsection{TRSB instability in the fixed amplitude limit}
\label{subsec:tri_supp_analytic}

We derive the TRS wall in the constant-amplitude, weak-strain limit and then determine
its lowest axial mode. This gives the linear instability
criterion and the leading spatial form of the emerging axial order.
 
Let $(\rho_R(\zeta),\beta_R(\zeta),0)$ be a stationary TRS solution
of the dimensionless equations. For a small  $\eta(\zeta)$ that
vanishes at the boundaries, expansion of
Eq.~\eqref{eq:tri_supp_density} up to quadratic order gives
\begin{equation}
\delta\mathcal F^{(2)}=\int d\zeta\,
\left[K_R(\eta')^2+U_\chi\eta^2\right],
\label{eq:tri_supp_quadratic}
\end{equation}
where $K_R={x_s^2\rho_R^2}/{4}$ and 
\begin{equation}
\begin{aligned}
U_\chi=g\rho_R^4+\frac32v\rho_R^6\cos3\beta_R-K_R(\beta_R')^2
+\frac{\rho_R^2}{2}h_\parallel(\zeta,\beta_R).
\end{aligned}
\label{eq:tri_supp_Uchi}
\end{equation}
 
The associated fluctuation operator for $\eta$ is
\begin{equation}
\begin{aligned}
\mathcal L_\chi&=-\partial_\zeta(K_R\partial_\zeta)+U_\chi.\\
\end{aligned}
\label{eq:tri_supp_operator}
\end{equation}

The smallest eigenvalue of this operator will determine whether there is an instability towards forming TRSB states. 
\begin{equation}
\lambda_{\min}=\inf_{\eta\ne0}
\frac{\int d\zeta\,[K_R(\eta')^2+U_\chi\eta^2]}
{\int d\zeta\,\eta^2}.
\label{eq:tri_supp_operator_lam}
\end{equation}
 
Thus negative numerator in Eq.~\eqref{eq:tri_supp_operator_lam} for any trial function implies that the TRS domain wall is unstable toward axial order. We next proceed to calculate the spectrum of this operator in the limit of fixed amplitude and weak strain limit, with $\rho(z)=\bar\rho$, with $\bar\rho^2\simeq-t/2$ for $t<0$. The fixed amplitude limit is well justified when the amplitude healing length $\xi_\rho = \sqrt{d/2 u \bar{\rho}^2}$ is much smaller than $L$.   

We introduce the dimensionless constants
\begin{equation}
K=\frac{x_s^2\bar\rho^2}{4},\qquad
G=g\bar\rho^4,\qquad V=v\bar\rho^6.
\label{eq:tri_dimless_KGV}
\end{equation}
The strain is taken to select the asymptotic domains while giving a
negligible correction to the wall-core equation. After subtracting the
constant isotropic condensation energy $t\bar\rho^2+\bar\rho^4$,
the angular free-energy density becomes
\begin{equation}
\begin{aligned}
\widetilde f_{\mathrm{ang}}={}&K\left[(\eta')^2
+\cos^2\eta(\beta')^2\right]
\\ &+G\sin^2\eta+V(1-\cos^3\eta\cos3\beta).
\end{aligned}
\label{eq:tri_dimless_angular}
\end{equation}
where primes denote derivatives with respect to $\zeta$.

Choose neighboring nematic domains with
$\beta_R(-\infty)=0$ and $\beta_R(+\infty)=2\pi/3$.
The reversed wall follows by reversing $\zeta$ and has the same
instability criterion. In the TRS sector $\eta=0$, the density is
$\widetilde f_R=K(\beta_R')^2+V(1-\cos3\beta_R)$, and its
Euler--Lagrange equation is
\begin{equation}
2K\beta_R''=3V\sin3\beta_R.
\label{eq:tri_dimless_beta_eq}
\end{equation}
Multiplying by $\beta_R'$ and integrating once gives
\begin{equation}
K(\beta_R')^2=V(1-\cos3\beta_R),
\label{eq:tri_dimless_first_integral}
\end{equation}
where the integration constant vanishes because the gradient and
anisotropy energy both vanish in the bulk. For the increasing solution,
$\beta_R'=\sqrt{2V/K}\sin(3\beta_R/2)$.
Integration yields
\begin{equation}
\begin{aligned}
\beta_R(\zeta)&=\frac43\arctan\exp\!\left(
\frac{\zeta-\zeta_0}{\ell_\beta}\right),\\
\ell_\beta&=\sqrt{\frac{2K}{9V}}
=\frac{x_s}{\sqrt{18v}\,\bar\rho^2}.
\end{aligned}
\label{eq:tri_dimless_kink}
\end{equation}
Here $\ell_\beta$ is the dimensionless wall width and $\zeta_0$ is
its center, which is free in the zero-field equation and is selected
by the weak imposed strain. Writing $x=(\zeta-\zeta_0)/\ell_\beta$,
we obtain the identities
\begin{equation}
\begin{aligned}
\beta_R'=\frac{2}{3\ell_\beta}\operatorname{sech}x, \quad 
\cos3\beta_R=1-2\operatorname{sech}^2x.
\end{aligned}
\label{eq:tri_dimless_identities}
\end{equation}

For this background, neglecting strain and amplitude variation gives
the axial fluctuation operator $\mathcal{L}_\chi$ 
\begin{equation}
\begin{aligned}
\mathcal L_\chi 
  = -K\partial_\zeta^2+G+\frac32V-5V\operatorname{sech}^2x.
\end{aligned}
\label{eq:tri_dimless_operator}
\end{equation}
Since $K/\ell_\beta^2=9V/2$, this operator can be
factorized as
\begin{equation}
\begin{aligned}
\mathcal L_\chi&=\frac{9V}{2}\mathcal A^\dagger\mathcal A+G-\frac V2,
\end{aligned}
\label{eq:tri_dimless_factorization}
\end{equation}
with $\mathcal A=\partial_x+ 2\tanh x/3.$
 
For functions vanishing at infinity,
$\int dx\,u^*\mathcal A^\dagger\mathcal A u
=\int dx\,|\mathcal A u|^2\geq0$.
Furthermore, $\mathcal A u_0=0$ has the normalizable solution
\begin{equation}
u_0(x)\propto\operatorname{sech}^{2/3}x.
\label{eq:tri_dimless_mode}
\end{equation}
It attains the lower bound, so the lowest eigenvalue is exactly
\begin{equation}
\begin{aligned}
\lambda_{\min}&=G-\frac V2
=\bar\rho^4\left(g-\frac{v\bar\rho^2}{2}\right),
\end{aligned}
\label{eq:tri_dimless_eigenvalue}
\end{equation}
within the constant-amplitude, weak-strain approximation.

The same result gives an explicit variational comparison. Keeping
$\beta=\beta_R$ and $\rho=\bar\rho$, choose the small localized
deformation $\eta(\zeta)=\epsilon\operatorname{sech}^{2/3}x$.
The common phase adjusts through the zero-current relation, and
$\eta(\pm\infty)=0$ preserves the TRS bulk domains. The energy change at lowest order of $\epsilon$ is
\begin{equation}
\begin{aligned}
\Delta\mathcal F={}&\epsilon^2\bar\rho^4
\left(g-\frac{v\bar\rho^2}{2}\right)\int d\zeta\,\operatorname{sech}^{4/3}x.
\end{aligned}
\label{eq:tri_dimless_variation}
\end{equation}
Thus a sufficiently small nonzero deformation in the axial channel lowers the
energy when
\begin{equation}
{g<\frac{v\bar\rho^2}{2}.}
\label{eq:tri_dimless_criterion}
\end{equation}
Here $g$, $v$, and $\bar\rho$ are all dimensionless, with the
normalization defined above.

\bibliography{references}

\end{document}